\documentclass[fp]{jpsj3}
\usepackage{subfigure}

\title{Defect-Mediated Aggregation and Motility-Induced Phase Separation in Self-Propelled Lattice-Gas Active XY Model}

\author{Shun Inoue\thanks{shun007171@gmail.com}%
and Satoshi Yukawa\thanks{yukawa@ess.sci.osaka-u.ac.jp}}
\inst{Department of Earth and Space Science,\\%
Graduate School of Science, The University of Osaka,\\%
Toyonaka, Osaka 560-0043, Japan}

\abst{We propose a ``self-propelled lattice-gas active XY (SPLG-AXY) model'' that incorporates key elements of both the classical XY model and the Vicsek model to study the role of topological defects in active matter systems.
This model features self-propelled particles with XY spin degrees of freedom on a lattice and introduces a self-propulsion parameter controlling the directional bias of particle motion.
Using numerical simulations, we demonstrate that self-propulsion induces motility-induced phase separation (MIPS), where particles aggregate into clusters around topological defects with positive vortex charge. In contrast, negative charge defects tend to dissipate. 
We analyze the evolution of these clusters and show that their growth follows a two-stage exponential relaxation process, with characteristic time scaling as $\tau \sim L^{3}$ with the system size $L$, reminiscent of first-order phase separation in equilibrium systems. 
Our results highlight the important role of topological defects in phase separations and clustering behavior in active systems, bridging nonequilibrium dynamics and equilibrium theory.}

\begin{document}
\maketitle

\section{Introduction}
\label{sec:introduction}

An active matter system, a collection of self-propelled particles that exhibit complex collective behavior, has recently attracted significant attention in statistical physics and nonequilibrium dynamics.\cite{Marchetti2013, Bechinger2016, Toner2024} Natural phenomena such as bird flocks \cite{Reynolds1987, Shimoyama1996}, fish schools \cite{niwaNewtonianDynamicalApproach1996, ItoUchida2022, itoEmergenceGiantRotating2022}, and bacterial colonies \cite{kumar2010,nishiguchi2025} demonstrate cooperative behaviors that are difficult to explain within a theoretical framework for equilibrium systems. Recent experimental studies have revealed an intrinsic connection between the collective behavior of active matter and topological defects for the neural stem cells aggregating and dispersing around topological defects \cite{kawaguchi2017topological} and the formation of vortex structures in biomolecular motor systems \cite{sumino2012large}. 

A representative active matter model is the Vicsek model \cite{vicsek1995novel}, in which self-propelled particles align their directions through local interactions, exhibiting flocking behavior. However, the feature of topological defects in the Vicsek model remains unclear.
In contrast, the two-dimensional classical XY model \cite{stanley1968dependence,tobochnik1979monte}, which serves as a theoretical foundation for the Vicsek model, has long established the significance of topological defects, that is, vortex defects in the low-temperature phase, most notably through the Kosterlitz-Thouless transition \cite{kosterlitz1973,kosterlitz1974critical,berezinskii1971destruction,berezinskii1972destruction}. The properties of the topological defects in the two-dimensional classical XY model, such as the mechanism of creating a pair of topological defects and the sustenance of the quasi-long-range order in spins, are well elucidated theoretically. \cite{Nelson2002}

This article proposes a novel ``self-propelled lattice-gas active XY (SPLG-AXY) model'' that integrates the key features of the two-dimensional classical XY and Vicsek models to investigate the relation between the topological defects and particle clustering dynamics. 
The present model combines the particle interactions inherited from the classical XY model with the self-propulsion-driven particle motion from the nonequilibrium nature of the Vicsek model. 
This framework is designed to advance our understanding of the role of topological defects in active matter. \cite{shankar2022}
In particular, we provide a detailed analysis of how the prominent topological defects found in the classical XY model at low temperatures are modified by introducing self-propulsion and how these modifications are related to motility-induced phase separation (MIPS), which is observed in the active matter systems frequently \cite{Cates2015}, and particle aggregation phenomena.

The present model evokes a lattice active matter system with Ising-type internal degrees of freedom \cite{ST13,kourbane-houssene2018, AadachiNakano2024, NakanoAdachi2024}; 
Unlike those models, the internal degrees of freedom of the present model are continuous spin variables.
The present model also resembles the continuous version of the Vicsek model, namely the Toner-Tu model. \cite{toner1995long, tonerFlocksHerdsSchools1998, Toner2012, Toner2024} 
In contrast, the present model is defined on a square lattice and allows unoccupied lattice sites, which distinguishes it from the Toner-Tu model. 
Furthermore, the Toner-Tu and Vicsek models exhibit a traveling-wave phase in some parameter region, and the feature is absent in the present model as discussed in Sec.~\ref{sec_snapshot}. In this sense, the behavior of the present model is closer to that of active model B plus (AMB+) \cite{NFTWTC17, TNC2018}.
It should be noted that there is another ``active XY model'' proposed by Haldar \textit{et al.} \cite{haldarMobilityinducedOrderActive2023,haldarActiveXYModel2023}
The two models are conceptually entirely different: In their model, spin variables are fixed at lattice sites, and walkers have neither spin variables nor a self-propulsion force.

The organization of this paper is as follows: In Section \ref{modelsimulation}, we construct the model and describe the simulation methods in detail. We then present simulation results that quantitatively demonstrate the emergence of vortex defects and cluster formation, and their scaling behavior as functions of the self-propulsion and particle density parameters in Section \ref{results}. 
Finally, we discuss the physical implications of our findings and the insights offered by the SPLG-AXY model into phase transitions in nonequilibrium systems.

\section{Model and Simulation}
\label{modelsimulation}

In this section, we define the model used in the present study. A brief sketch of the model is as follows: The active particles have a two-dimensional constant-length vector as internal degrees of freedom, representing the head direction. Hereafter, we call the vector an XY spin. The active particle lives on the lattice site. The model is then extended by allowing each particle to move on lattice sites according to the spin direction.

\subsection{Self-Propelled Lattice-Gas Active XY Model}
\label{subsec:activexymodel}

We construct a simple model in which the particle having an XY spin moves on lattice sites, which we refer to as the SPLG-AXY model. In this study, we take a two-dimensional square lattice of size $L \times L$.

\begin{figure}[htbp]
\centering
\includegraphics[width=0.25\textheight]{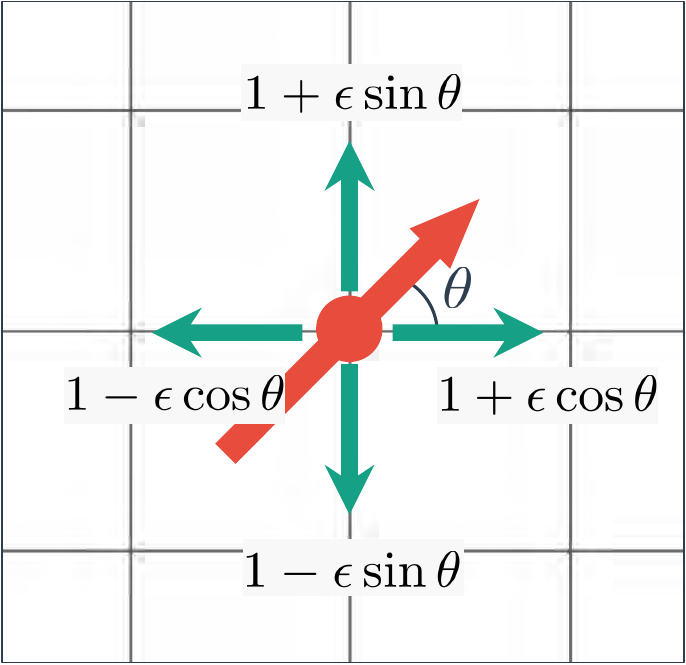}
\caption{(Color online) Schematic picture of particle movement rates and its orientation. The movement rate depends on the self-propulsion parameter $\epsilon$ and the spin orientation.}
\label{fig:activexy}
\end{figure}
Particles at each lattice site can move probabilistically in four directions: up, down, left, and right. The XY spin orientation is defined by an angle $\theta$ as in Fig.~\ref{fig:activexy}, where $\theta=0$ corresponds to the rightward direction, and the angle increases counterclockwise. 
A weight parameter $\epsilon$ ($0 \leq \epsilon \leq 1$), representing an intensity of the self-propulsion, is introduced such that the particle movement rate depends on $\epsilon$.
For the general $\theta$ case, probability weights are shown in Fig.~\ref{fig:activexy}; 
For example, suppose a particle is oriented to the right ($\theta=0$) and $\epsilon = 1$. In that case, the probability of choosing the right lattice site as a moving site is $1/2$, and the probability of selecting either the upper or lower lattice sites is $1/4$. The probability of choosing the left lattice site is $0$. 
In the case of $\epsilon=0$, the site is chosen with equal probability $1/4$ in all four directions, independent of its orientation.

Furthermore, we assume an exclusion principle corresponding to a repulsive interaction, whereby multiple particles cannot occupy the same lattice site. A particle first selects its direction of motion probabilistically, as mentioned above. However, if another particle already occupies the target lattice site, it remains at its original site. No further attempts to move to an alternative site are made during the same step. 

The periodic boundary condition is imposed for the particle motion. The total number of particles is denoted by $n$, and the particle density is defined as $\rho = n \slash L^2$.

\subsection{Interactions of Spins (Classical XY Model)}

The interaction between XY spins is the same as that of the classical ferromagnetic XY model. However, in the SPLG-AXY model, since particles can move between lattice sites, there are cases where a lattice site may be unoccupied. In such instances, the interaction is extended to yield zero energy. 
Specifically, we label the lattice sites with consecutive numbers \( i = 1,2,\dots,L^2 \) and introduce an occupancy variable $X_{i}$ for each site \( i \):
\[
  X_i =
  \begin{cases}
    1 & \text{if a particle exists at the site $i$,} \\
    0 & \text{if the site $i$ is empty.}
  \end{cases}
\]
For a occupied site \( i \), an XY spin with an angle \(\theta_i\) is assigned as
\[
  \boldsymbol{s}_i = (\cos\theta_i, \sin\theta_i).
\]
When \( X_i = 0 \), the spin is not defined and does not contribute to the interaction.
Then, the interaction energy is given by
\begin{equation}
  E = - J \sum_{\langle i,j\rangle} X_i\,X_j \,\cos(\theta_i - \theta_j)
  \label{eq:active_xy_interaction}
\end{equation}
where \(\langle i, j\rangle\) denotes pairs of nearest-neighbor sites, and \(J (>0) \) is the coupling constant. In this study, we set \(J=1\). 
This expression represents the ``alignment'' interaction of the active particles.

When particles occupy every lattice site due to the exclusion rule at \(\rho=1\), the present model exactly reproduces the classical XY model, and the interaction energy reduces to
\[
  E = -J \sum_{\langle i,j\rangle} \cos(\theta_i - \theta_j).
\]
By introducing the occupancy variable \(X_i\), the SPLG-AXY model accounts 
for cases in which some lattice sites are empty.

Equation~\eqref{eq:active_xy_interaction} implies that the interaction of the classical ferromagnetic XY model acts only between sites where particles are present. Consequently, the SPLG-AXY model in this study incorporates the equilibrium characteristics of the classical XY model in addition to the nonequilibrium feature of self-propelled motion. 
This interaction can also be interpreted as an interaction in which particles attempt to align their orientations with their neighbors while moving, similar in spirit to the Vicsek model.

The system's temperature $T$ is also introduced, and the XY spins are simulated using the Metropolis Monte Carlo method. 
In this study, the temperature is fixed at $T=J/4$, corresponding to the low-temperature phase of the classical XY model with $\rho=1$.

\subsection{Detalied Update Algorithm}
\label{subsec:algorithm}

Here, we describe the specific algorithm for the SPLG-AXY model. First, a square lattice of size $L \times L$ is prepared, and $n$ particles are randomly placed on each lattice site under the exclusion rule. Each particle is assigned an XY spin orientation, which is randomly set as the initial condition, and a label identifying the particle. 

Time evolution consists of the following two procedures: The movement and orientation update processes. For the movement process, the set of labels of all particles is randomly shuffled, and each particle is sequentially allowed to attempt a move on the lattice according to the order of the shuffled labels. The Fisher-Yates shuffle algorithm \cite{fisher1963statistical, knuth1997} is employed to shuffle the labels.
The movement rates are already mentioned in Sec.~\ref{subsec:activexymodel}.
Moves to lattice sites already occupied by another particle are not permitted. Furthermore, if a move cannot be executed, the particle does not attempt to move again during the same time step. 
For the orientation update process, the labels of all particles are randomly shuffled once again, and each particle is sequentially allowed to attempt an update of its orientation according to the shuffled labels. The orientation update is performed based on the interactions of the classical XY model (Eq.~\eqref{eq:active_xy_interaction}), and the Metropolis Monte Carlo method is used during the update;
A new state $\theta_{i}$ of the orientation by a uniform random distribution is proposed. 
Then, we calculate the energy difference $\Delta E$ between the proposed and current states. Acceptance of the proposed state is based on the following criteria: If $\Delta E \le 0$, that is, the new state has lower or equal energy than the current state, we accept the new state. If $\Delta E > 0$, that is, the new state has higher energy than the current state, we accept the new state with probability $\exp (-\Delta E \slash k T)$, where $k$ is the Boltzmann constant and $T$ is the temperature. If the new state is accepted, it becomes the current state. If rejected, the system remains in the current state.

The above two movement and orientation processes constitute one Monte Carlo step (MCS) during the simulation. 
Once sufficient time has elapsed since the initial condition and the physical quantities have converged, the system is considered to have reached a steady state. It should be noted that due to the self-propelled nature of the particles in the SPLG-AXY model, the system is in a nonequilibrium steady state.
The computational performance of the SPLG-AXY model, compared with the Vicsek model and AMB+, is summarized in Appendix\ref{sec:compperf}.

\section{Results}
\label{results}

In this section, we summarize the results obtained using the SPLG-AXY model. All simulations are conducted at a temperature of $T=J\slash 4$, corresponding to the low-temperature phase of the classical XY model in which topological defects emerge.

\subsection{Snapshots}
\label{sec_snapshot}
\begin{figure}
	\centering\includegraphics[width=0.98\linewidth]{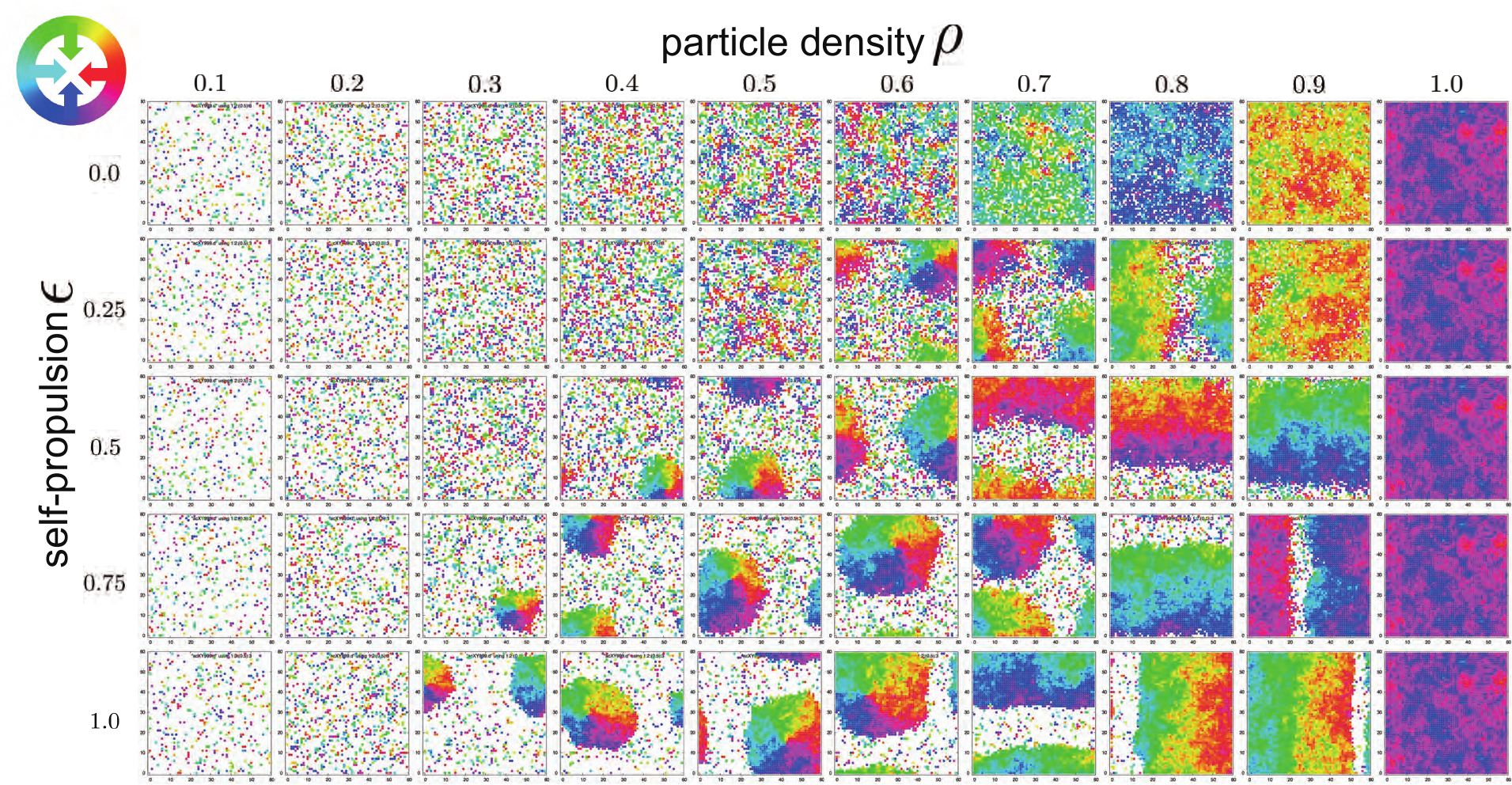}
	\caption{(Color online) Snapshots of the simulation for various values of $\rho$ and $\epsilon$. 
The density increases to the right, and the self-propulsion parameter increases downward in this grid layout. 
Particle orientation is indicated by hue, with the corresponding color wheel and directional reference shown in the upper-left corner. The white represents the empty sites.
	Increasing the self-propulsion parameter $\epsilon$ induces motility-induced phase separation (MIPS) in sufficiently high densities. These snapshots are obtained for the size $L=60$.}
\label{fig:snapshot-rhoeps}
\end{figure}
We show simulation snapshots obtained by varying the self-propulsion parameter $\epsilon$ and the particle density $\rho$ in Fig.~\ref{fig:snapshot-rhoeps} as a grid layout. The simulations are performed on a system of size $L=60$, and the snapshots are taken after 100\,000 MCSs. 
Particle orientation is indicated by hue, with the corresponding color wheel and directional reference shown in the figure.
It is observed that increasing the self-propulsion parameter $\epsilon$ triggers phase separation into two distinct phases: An aggregated region where the local density is $1$ and an unaggregated random walking region. 
This phase separation can be regarded as the motility-induced phase separation (MIPS).
Moreover, phase separation occurs even at lower values of $\epsilon$ when the density $\rho$ is sufficiently high.

\begin{figure}
	\centering
	\includegraphics[width=1.0\linewidth]{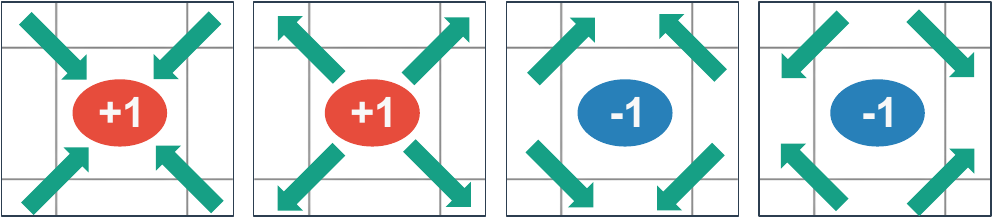}
	\caption{(Color online) Emerging topological defects. Due to the self-propulsion and the particle exclusion, the $m=+1$ topological defect shown on the leftmost figure tends to persist intuitively, whereas all instances of the $m=-1$ topological defects shown on the right two figures dissipate. Even among the $m=+1$ topological defects, the topological defect shown on the second left is less likely to serve as a nucleus for cluster formation, suggesting that the leftmost defect acts as the primary nucleus.}
	\label{fig_vortex}
\end{figure}

Figure~\ref{fig:snapshot-rhoeps} shows that the phase separation is thought to be induced by the combined effects of self-propulsion $\epsilon$ and exclusion. For the case of $\epsilon\ne 0$, clear phase separation is observed. 
For the case of $\epsilon=0$, however, there is no clear phase separation. In the middle densities, there is a disk-shaped cluster. In addition, this cluster has a single topological defect at the center, which is recognized as a singular color point of the cluster. This topological defect has an $m=+1$ vortex charge, which is defined as an integral of the orientation gradient $\nabla\theta$ on the closed loop (see also Fig.~\ref{fig_vortex}):
\begin{equation}
m = \dfrac{1}{2\pi} \oint_{\square} \nabla \theta \cdot  \mathrm{d}\boldsymbol{\ell}.
\label{eq_charge}
\end{equation}
The particle orientation near the topological defect shows alignment toward the center, as clearly shown in the hue-based visualization of orientation.
Thus, the topological defect is a radial type, not a tangential one. 
For higher densities, this cluster extends across the entire system length and merges with itself due to the periodic boundary conditions. In this situation, the cluster has no topological defect. 
This band structure observed in Fig.~\ref{fig:snapshot-rhoeps} does not exhibit a traveling wave motion. 
This transition will be discussed later.  
It is confirmed that the clusters grow in size as the density $\rho$ increases. Although Fig.~\ref{fig:snapshot-rhoeps} suggests that further increasing the self-propulsion parameter $\epsilon$ enhances the cluster size, the detailed dependence has not been investigated in the present study.

Based on the simulation results, phase separation through cluster formation is associated with topological defects. In the classical XY model, topological defects with $m=+1$ and $m=-1$ occur with equal probability. However, in the SPLG-AXY model, topological defects with $m=+1$ tend to persist preferentially. This is attributed to the fact that clusters form around $m=+1$ topological defects, while $m=-1$ topological defects are more prone to dissolution.
This fact is understood by schematic pictures in Fig.~\ref{fig_vortex}.
The left two schematic figures in Fig.~\ref{fig_vortex} have an $m=+1$ vortex charge, and the right two have $m=-1$ charge. This topological charge is easily calculated by the definition, Eq.~\eqref{eq_charge}.
In the leftmost case in Fig.~\ref{fig_vortex}, the particle blocks each other due to self-propulsion and particle exclusion. Therefore, this topological defect becomes the nucleus of the cluster. The particles can go away in the second case, even though the topological charge is also $m=+1$. However, in all cases of $m=-1$ charge, the particle cannot condensate near the defect. Thus, the cluster grows from the specific $m=+1$ topological defect. 
In the classical XY model, the spin orientation and the lattice direction are completely decoupled. Thus, the global spin rotation symmetry implies that the left two figures are in an identical configuration in Fig.~\ref{fig_vortex}. However, in the present case, the spin orientation is strongly coupled with the lattice direction due to the self-propulsion. Therefore, the two situations in the left figure are distinguished.

\begin{figure}
	\includegraphics[width=1.0\linewidth]{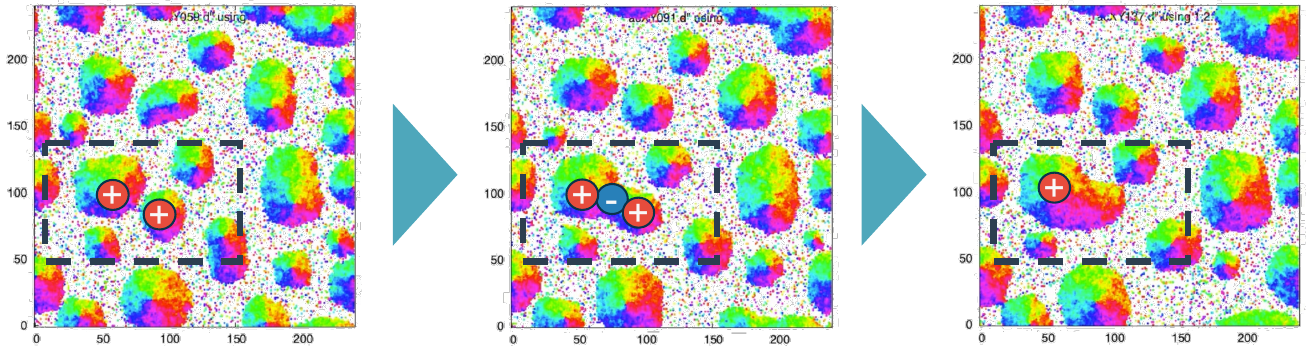}
	\caption{(Color online) Cluster merging process. The system size is taken to be $L=240$. 	$\pm$ signs represent topological defects with the vortex charge $m=\pm 1$.	As time increases from the left figure to the right one, we observe that two clusters with $m=+1$ vortex charge merge in a rectangular region shown by a dashed line. In the center, the two clusters are just touching. Then, a topological defect with $m=-1$ charge emerges at the contact point. In this situation, two $m=+1$ and one $m=-1$ topological defects exist in a single cluster. After relaxation, a pair of $m=+1$ and $m=-1$ topological defects annihilate, and a single $m=+1$ topological defect remains at the cluster's center.}
	\label{fig:cluster_grow}
\end{figure}
When an $m=+1$ topological defect forms, particles accumulate at the defect site, as explained with Fig.~\ref{fig_vortex}. As time evolves, $m=+1$ topological defects merge and grow into larger clusters. This growing process is shown in Fig.~\ref{fig:cluster_grow}; In the early stage, several clusters with $m=+1$ vortex charge emerge. As time passes, the clusters grow and occasionally come into contact. When the cluster contacts another cluster, the topological defect with $m=-1$ vortex charge emerges at the contact point. In this situation, two $m=+1$ and one $m=-1$ topological defects exist in a single cluster. After sufficient relaxation, a pair of $m=+1$ and $m=-1$ topological defects annihilate, and a single $m=+1$ topological defect remains at the cluster's center. Repeating this process, a single large cluster finally exists in the middle densities. 
A growing cluster extends over the system length and contacts itself due to the periodic boundary condition for higher densities. After sufficient relaxation, the cluster without the topological defect remains in this situation because the contacted cluster initially has a pair of $m=\pm 1$ topological defects (see also Fig.~\ref{fig:cluster_boundary}).

\subsection{Relationship between Total Vorticity $N$ and Self-Propulsion $\epsilon$ and Density $\rho$}
\label{subusubsec:Nandepsilon}

As shown in Fig.~\ref{fig:snapshot-rhoeps}, depending on the self-propulsion parameter $\epsilon$ and the density $\rho$, a topological defect remains even in a steady state. To quantify the remaining vortices in the system, we define the total number of vortex charges $N$ as the difference between the number of $m=+1$ defects, $N_{+1}$, and the number of $m=-1$ defects, $N_{-1}$. The total vorticity $N$ is then defined as
\[
N = N_{+1} - N_{-1}.
\]

\begin{figure}
\centering
\subfigure[$\epsilon = 0.0$]{\includegraphics[width=0.48\linewidth]{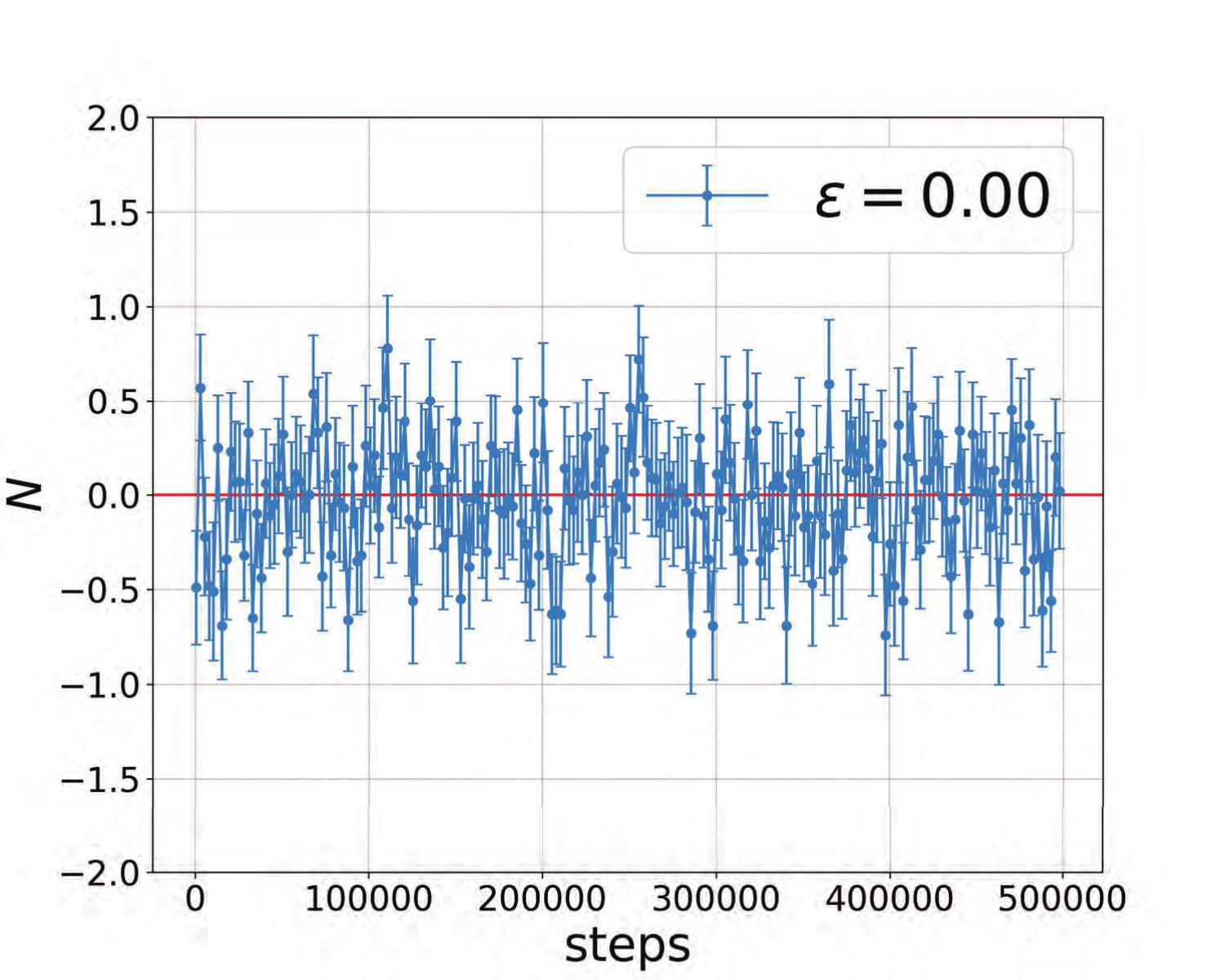}}
\subfigure[$\epsilon = 0.25$]{\includegraphics[width=0.48\linewidth]{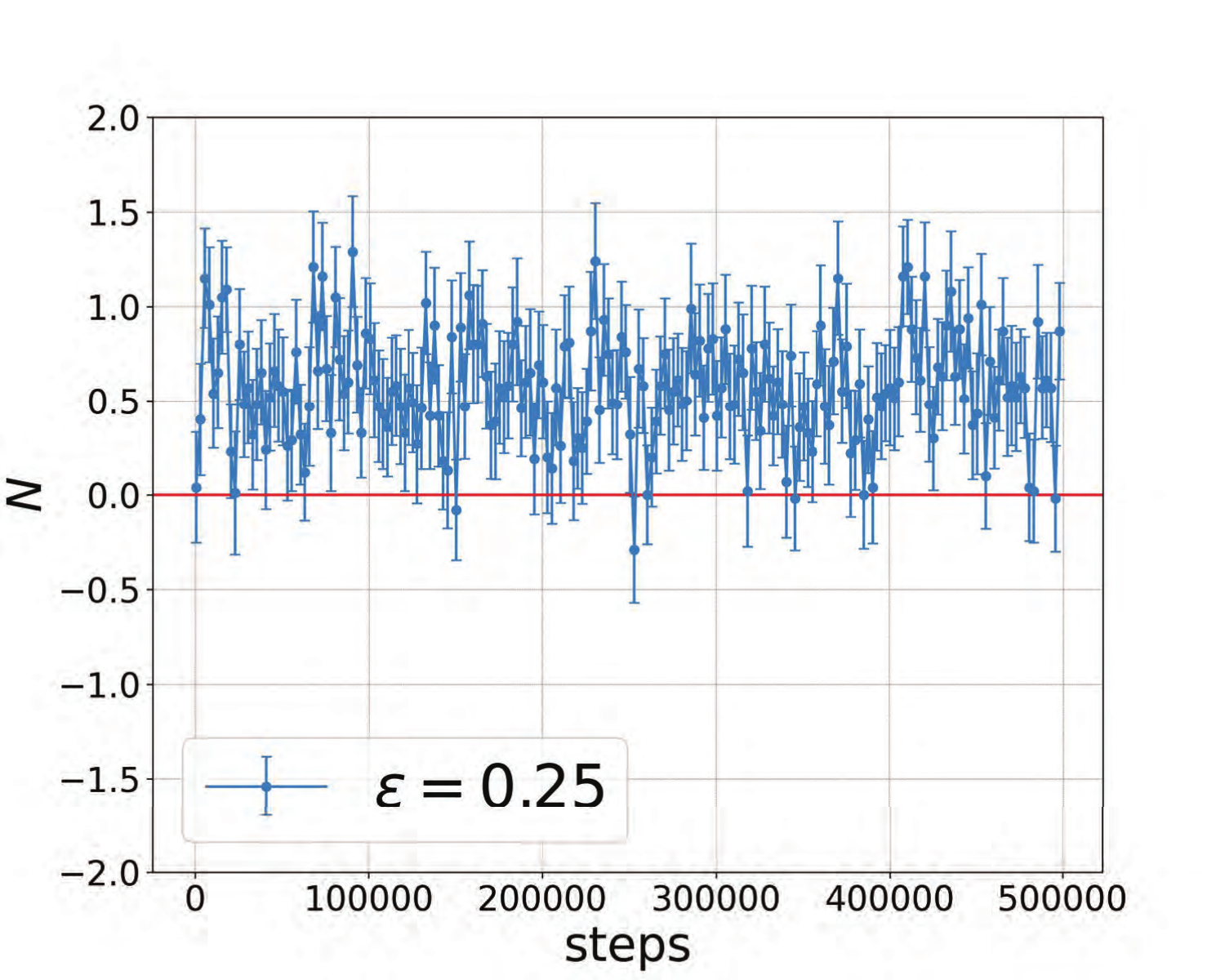}}
\subfigure[$\epsilon = 0.50$]{\includegraphics[width=0.48\linewidth]{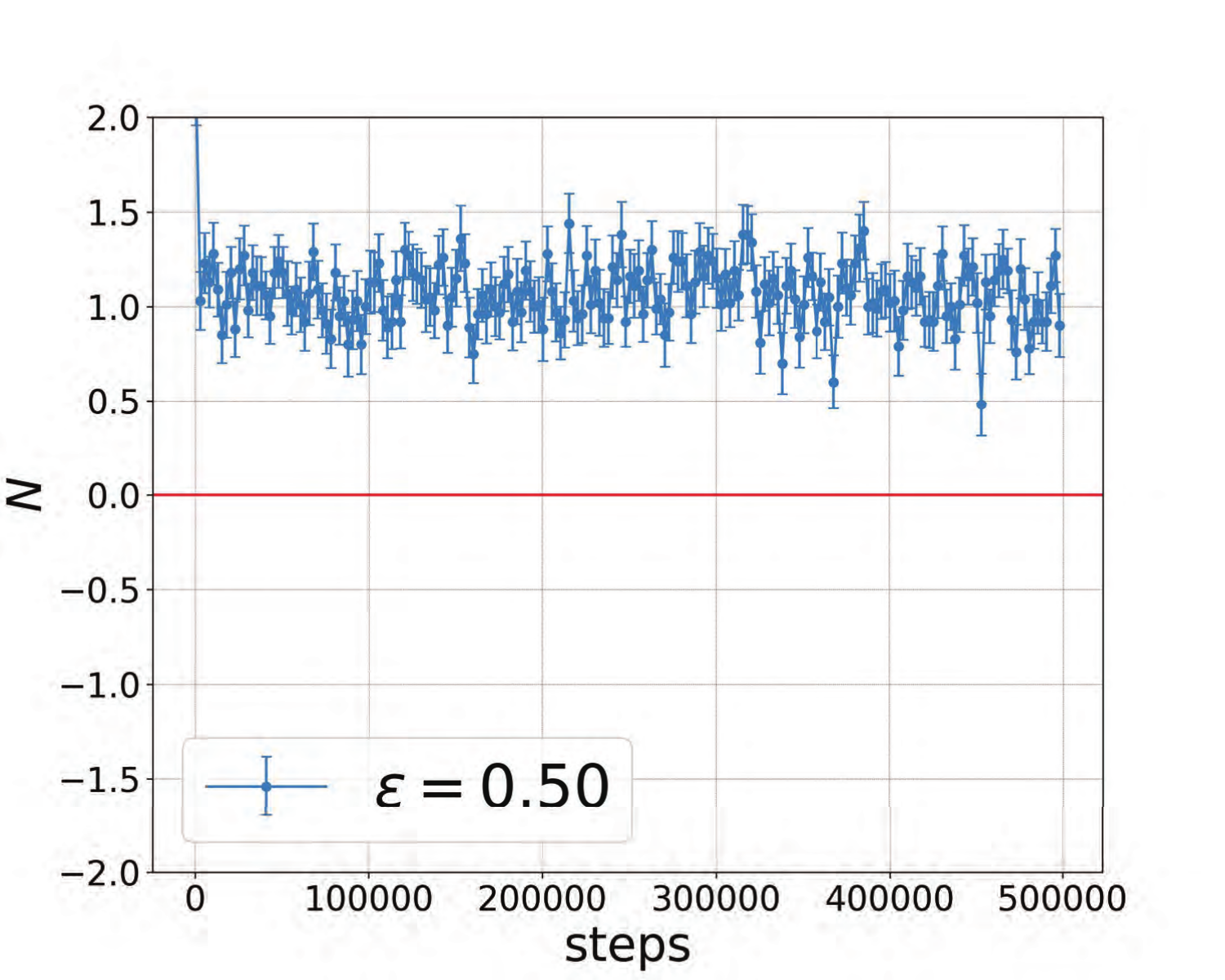}}
\subfigure[$\epsilon = 1.00$]{\includegraphics[width=0.48\linewidth]{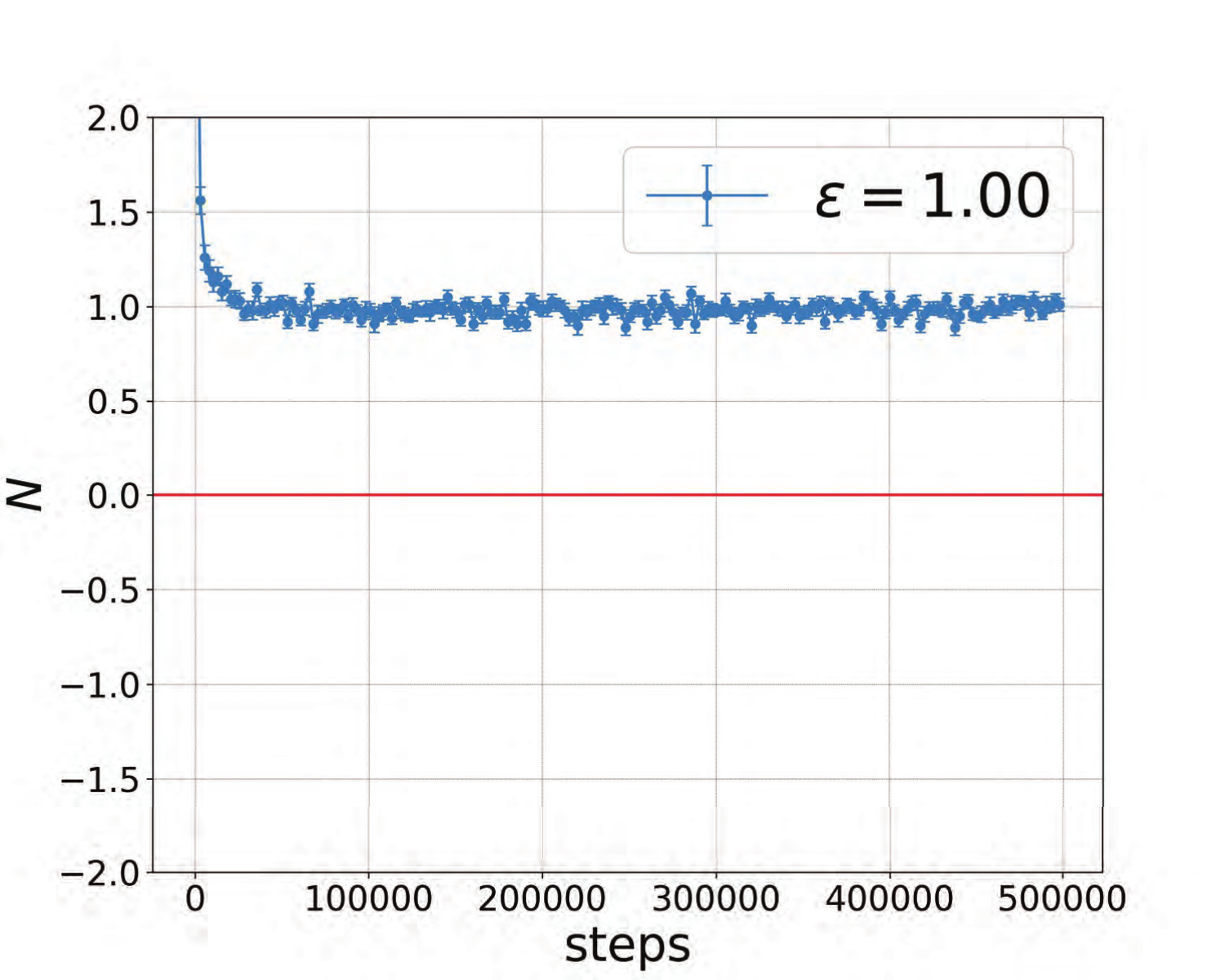}}
\caption{(Color online) Relationship between the self-propulsion parameter $\epsilon$ and the total vorticity $N$ at $\rho = 0.5$ and $L=60$.
The total vorticity is calculated every 2\,500 MCSs after discarding the initial 500 MCSs for initial relaxation.
The initial configuration is random, and the average vorticity is near zero. 
The horizontal guideline is indicated at $N = 0$.
The figure shows an average of over 100 samples.
For $\epsilon=0.0$, corresponding to a randomly moving XY model, the effect of self-propulsion is absent. Thus, the total vorticity converges to zero. As $\epsilon$ increases, the influence of the self-propulsion becomes apparent. Then, the total vorticity fluctuates around $N=1$. This total vorticity suggests that a single, large, disk-shaped cluster exists.}
  \label{fig:epsilonvortex}
\end{figure}
Figure~\ref{fig:epsilonvortex} shows the relationship between the self-propulsion parameter $\epsilon$ and the total vorticity $N$ when the particle density is fixed at $\rho = 0.5$ and the system size is $L=60$. 
The initial configuration is random, and the average vorticity is near zero. 
In the classical XY model, both $m=+1$ and $m=-1$ vortex defects occur with equal probability. For $\epsilon=0$, which corresponds to a randomly moving XY model, vortex defects appear with equal likelihood. Thus, the total vorticity converges to zero. 
As $\epsilon$ increases, the symmetry between $m=+1$ and $m=-1$ defects is broken by the self-propulsion. 
The total vorticity then fluctuates around $N=1$. This total vorticity shows quantitatively that $m=+1$ vortex defects tend to persist.

\begin{figure}
\centering
\subfigure[$\rho = 0.2$]{%
   \includegraphics[width=0.48\linewidth]{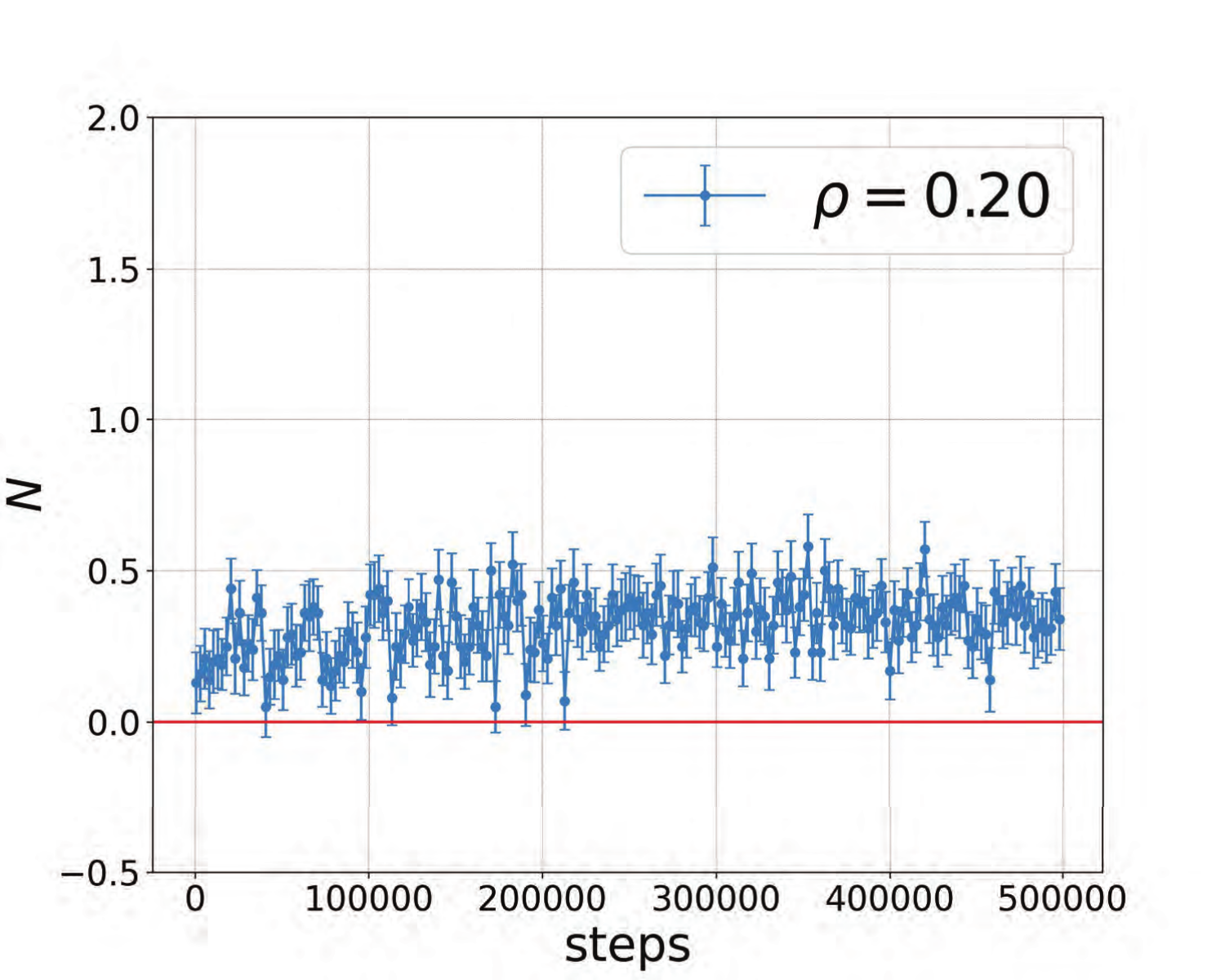}%
}
\subfigure[$\rho = 0.4$]{%
   \includegraphics[width=0.48\linewidth]{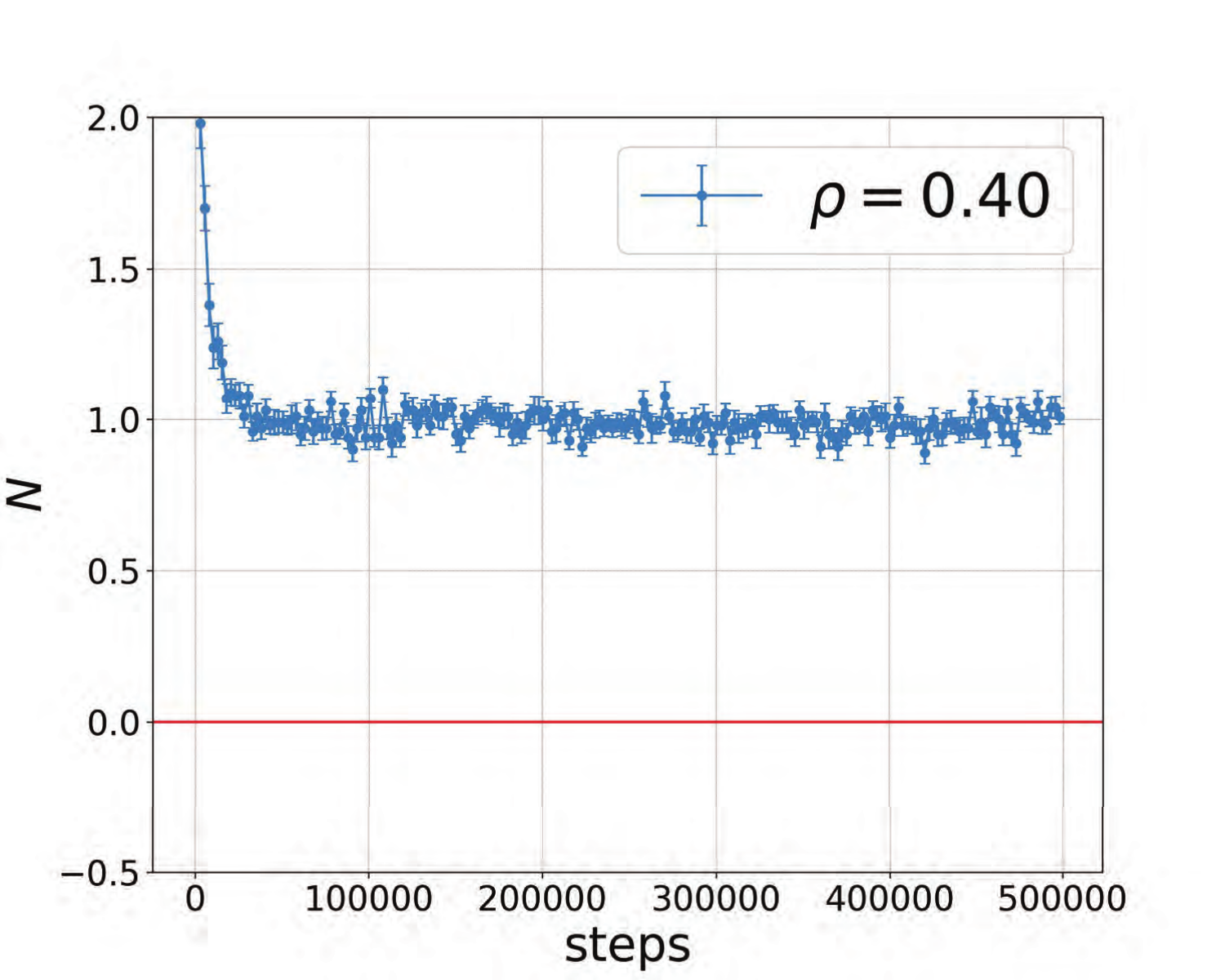}%
}
\subfigure[$\rho = 0.6$]{%
   \includegraphics[width=0.48\linewidth]{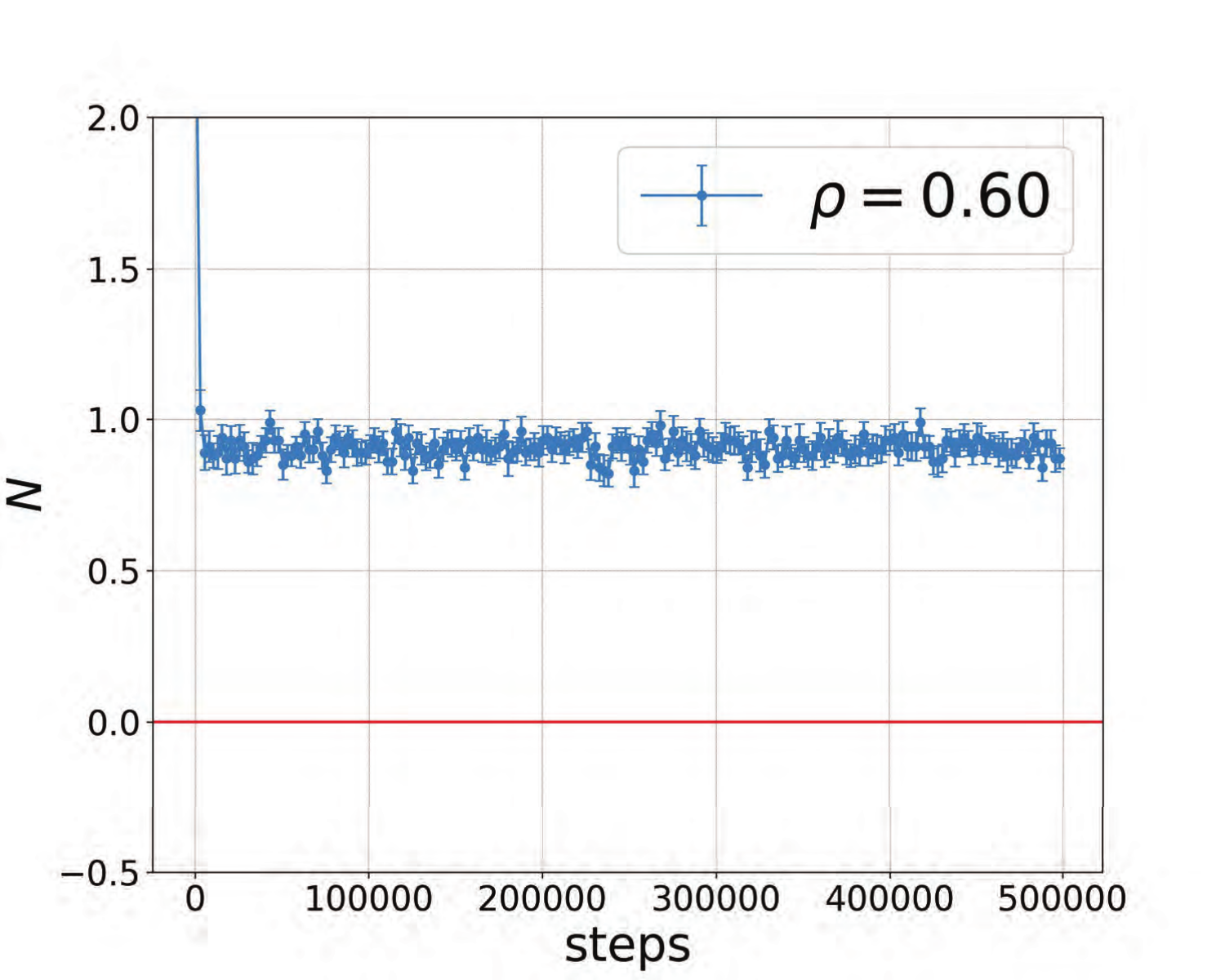}%
}
\subfigure[$\rho = 0.8$]{%
   \includegraphics[width=0.48\linewidth]{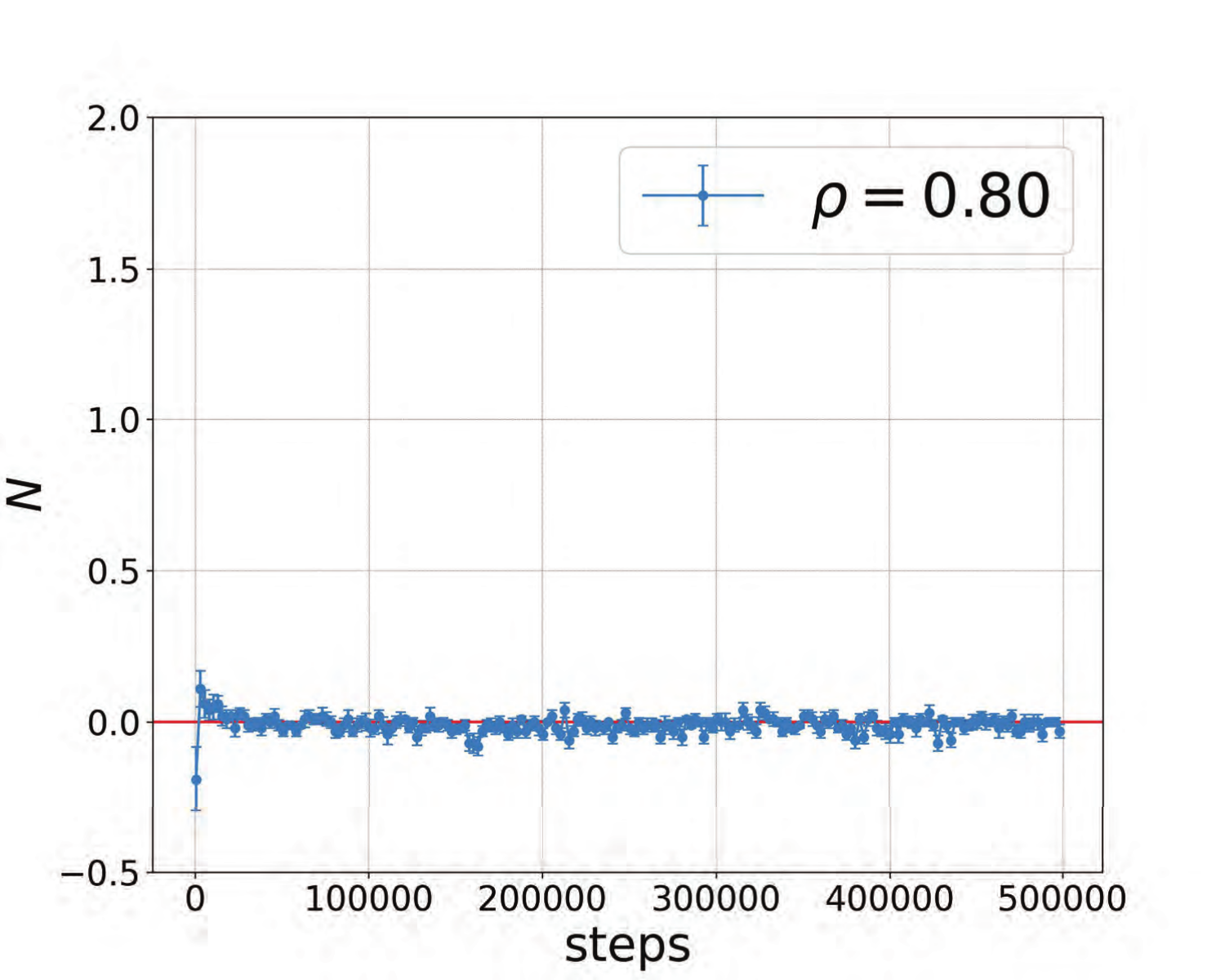}%
}
  \caption{(Color online) Relationship between the density $\rho$ and the total vorticity $N$ with the self-propulsion $\epsilon=1$ and $L=60$. The total vorticity is calculated every 2\,500 MCSs after discarding the initial 500 MCSs for initial relaxation. The initial configuration is random, and the average vorticity is near zero. The horizontal guideline is indicated at $N = 0$. The figure shows an average of over 100 samples. At $\rho=0$ and $\rho=1$, the total vorticity is exactly $N = 0$. As the particle number increases from $\rho=0$, the vorticity $N$ increases, but as $\rho$ approaches one, $N$ falls back to zero. This fact shows that the total vorticity peaks at an intermediate density. 
}
\label{fig:rhovortex}
\end{figure}
Next, we investigate the relationship between the particle density $\rho$ and the total vorticity $N$ in the high self-propulsion regime. Figure~\ref{fig:rhovortex} shows the dependence of the total vorticity $N$ on $\rho$ when the self-propulsion parameter is fixed at $\epsilon = 1$ and the system size $L=60$. In this case, the average vorticity at the initial condition is also near zero. At $\rho=0$, where no particles are present, and $\rho=1$, corresponding to the classical XY model, the vorticity is exactly $N=0$. Thus, as the particle number increases from $\rho=0$, the vorticity $N$ grows, but as $\rho$ approaches one, $N$ decreases back to zero. In other words, $N$ exhibits a peak at an intermediate density.

\begin{figure}
	\centering
	\includegraphics[width=1.0\linewidth]{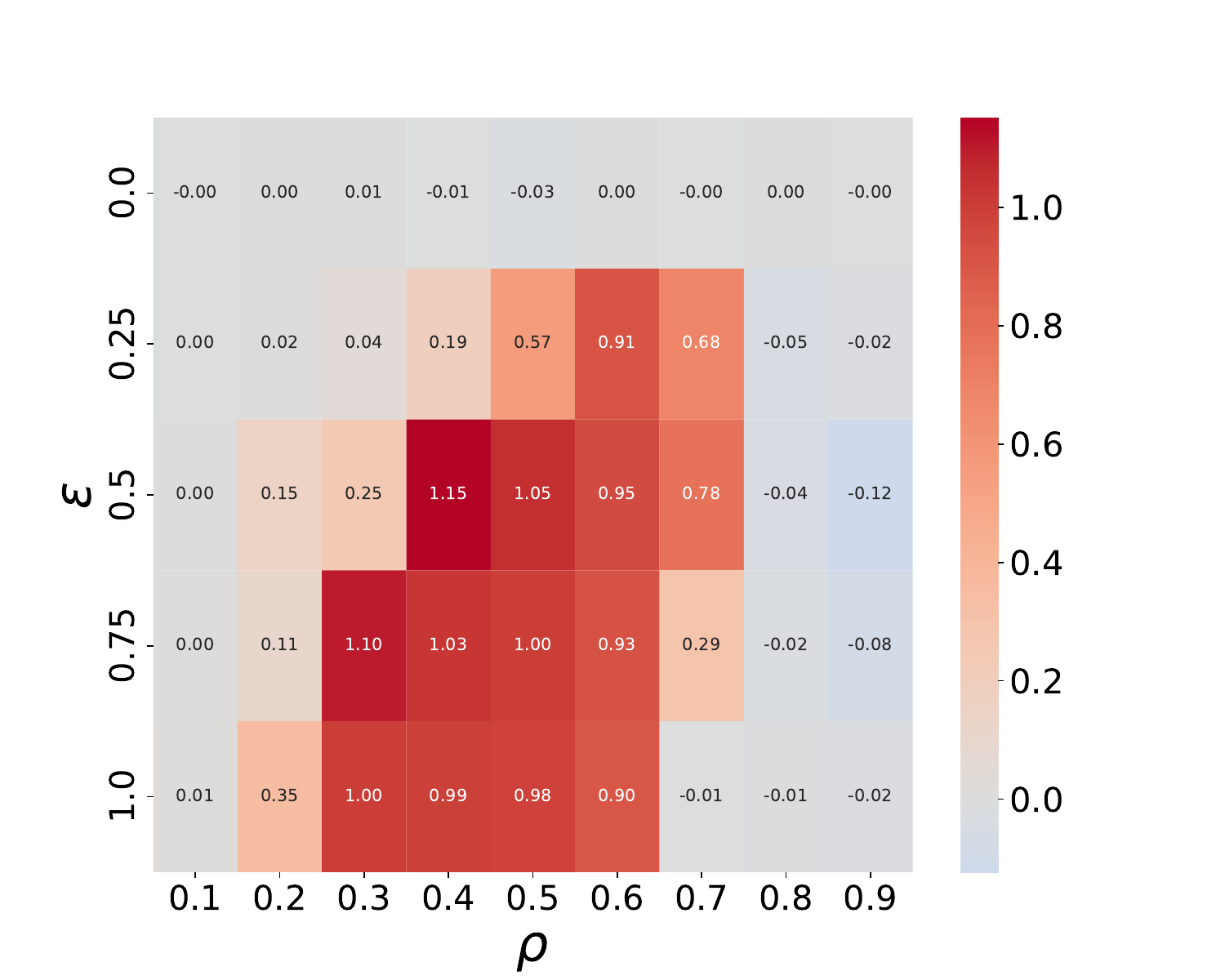}
	\caption{(Color online) 	Heatmap of the total vorticity $N$ on the $\epsilon-\rho$ plane. The heatmap is constructed from the data of the average of over $100$ samples of the total vorticity calculated over MCSs from 250\,000 to 500\,000. The system size is taken to be $L=60$.
}
\label{fig_heatmap}
\end{figure}
Based on the above results, we construct a heatmap that visualizes the dependence of the total vorticity $N$ on the self-propulsion parameter $\epsilon$ and the particle density $\rho$. The result is shown in Fig.~\ref{fig_heatmap}. The self-propulsion parameter is varied from $\epsilon=0$ to $1$ in increments of $0.25$, and the particle density is varied from $\rho=0.1$ to $0.9$ in increments of $0.1$. The average vorticity $N$ over $100$ samples is calculated over the time interval from 250\,000 to 500\,000 MCSs after the system reaches a steady state, and the resulting values are colored according to the value on the corresponding $\rho$ and $\epsilon$.

\begin{figure}
\centering
\includegraphics[width=1.0\linewidth]{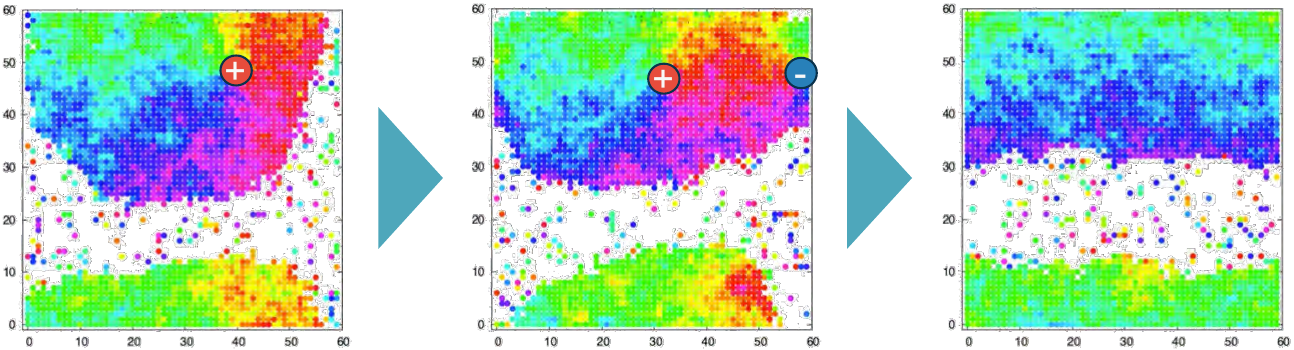}
\caption{(Color online) The effect of the periodic boundary on the cluster merging process. The system size is taken to be $L=60$. $\pm$ signs represent topological defects with the vortex charge $m=\pm 1$. As time increases from the left figure to the right one, we observe a single large cluster that connects itself across the boundary. Then, the $m=-1$ defect emerges at the connected point. After relaxation, the two topological defects annihilate.}
\label{fig:cluster_boundary}
\end{figure}
When the self-propulsion $\epsilon$ is small, the occurrences of $m=+1$ and $m=-1$ vortices tend to become nearly equal, and therefore the average vorticity approaches zero. As $\epsilon$ increases, the positive vorticity dominates. 
However, increasing $\epsilon$ does not lead to a positive vorticity $N$ at high densities.
This nearly zero total vorticity is because, in high-density regimes with strong self-propulsion, the system eventually forms one large cluster that does not contain a topological defect with vorticity $m=+1$. This cluster formation is an effect of the periodic boundary condition described in Sec.~\ref{sec_snapshot} and Fig.~\ref{fig:cluster_boundary}.

We can estimate the crossover density as the following simple argument:
Assuming that the radius of the disk-shaped cluster spanning the boundaries is approximately $r\simeq L/2$, the corresponding particle density can be estimated as 
\[
\dfrac{\pi}{L^2} \left(\dfrac{L}{2} \right)^{2}
= \frac{\pi}{4} \simeq 0.78.
\]
This density can be regarded as the crossover density at which the vorticity switches. However, in practice, the transition appears at a slightly lower density due to fluctuations and contributions from the unaggregated random walking particles.

Comparing Fig.~\ref{fig:snapshot-rhoeps} and Fig.~\ref{fig_heatmap}, we find a strong correlation between the total vorticity and the cluster formation of the active particle. In addition, in Fig.~\ref{fig_heatmap}, we observe that the peak of the total vorticity shifts toward higher densities as the self-propulsion $\epsilon$ decreases. 
This behavior may be related to the clearness of the cluster boundary at high density, which becomes more distinct or less fluctuates as the self-propulsion parameter $\epsilon$ increases.

\subsection{Quantitative Analysis of Cluster Formation during the Aggregation Process}

In the SPLG-AXY model, cluster formation and growth are observed due to the accumulation of particles. Many examples of aggregation phenomena driven by motility have been reported in active matter systems, and understanding these dynamics is crucial. In the following, we further investigate the dynamics of cluster growth under periodic boundary conditions in the high self-propulsion regime, $\epsilon=1$. 

\subsubsection{Snapshots for Different System Size}

\begin{figure}
	\centering
	\includegraphics[width=1.0\linewidth]{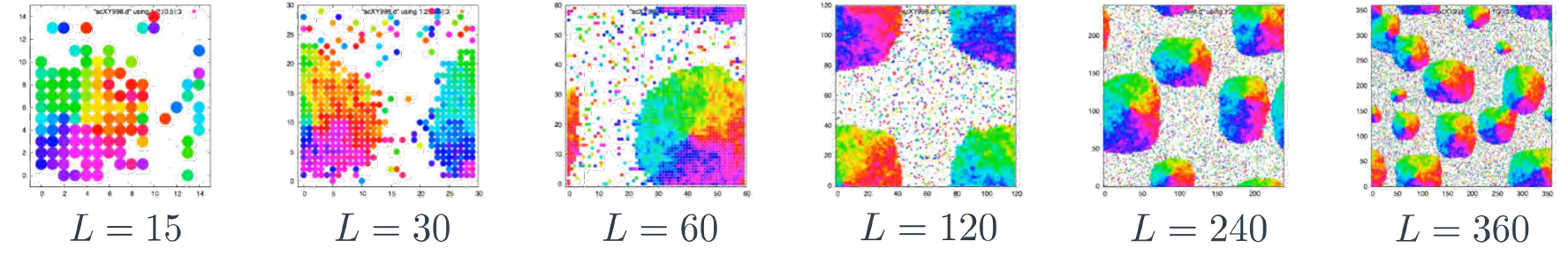}
	\caption{(Color online) Snapshots after the same elapsed time (100\,000 MCSs) with $\epsilon=1$ for different system sizes $L$ from the random initial configuration. In larger systems, the aggregation of clusters takes much longer MCSs.}
	\label{fig:size_cluster}
\end{figure}
In the previous section, we have demonstrated that clusters form through particle aggregation around topological defects. However, whether the growth process depends on the system size has not been examined. Thus, this section examines the dependence of the system size on the cluster growth process. Figure~\ref{fig:size_cluster} shows snapshots of the system for various system sizes $L=15, 30, 60, 120, 240,$ and $360$ after 100\,000 MCSs relaxation steps. For relatively large systems such as $L=240$ and $L=360$, the clusters have not yet fully merged into a single entity, and several intermediate-sized clusters are observed at this time step. 
However, it will be confirmed with further time evolution that they eventually coalesce into a single giant cluster.  
The snapshots shown in Fig.~\ref{fig:size_cluster} represent the existence of the relaxation time that depends on the system size.

These observations suggest that a single cluster eventually forms even as the system size increases. Therefore, the intrinsic characteristic of the SPLG-AXY model, in which particles predominantly aggregate via $m=+1$ topological defects leading to phase separation, is not merely a finite-size effect but is likely to occur even in large systems. Even if the initial state is random, the system converges to a single giant cluster after a sufficiently long relaxation time. Next, we will quantitatively analyze the cluster size distribution and the dynamics of cluster growth to further characterize the cluster formation and aggregation process.

\subsubsection{Cluster Size Distribution}
\label{subsec:cluster_distribution}

Since it has been confirmed that the system eventually forms a single giant cluster in the steady state, we now examine how the relaxation process varies with system size. Given that the nucleation of cluster formation is associated with $m=+1$ topological defects, the minimum cluster size may remain constant regardless of the system size.

\begin{figure}
\centering
\subfigure[$L=60$]{\includegraphics[width=0.47\linewidth]{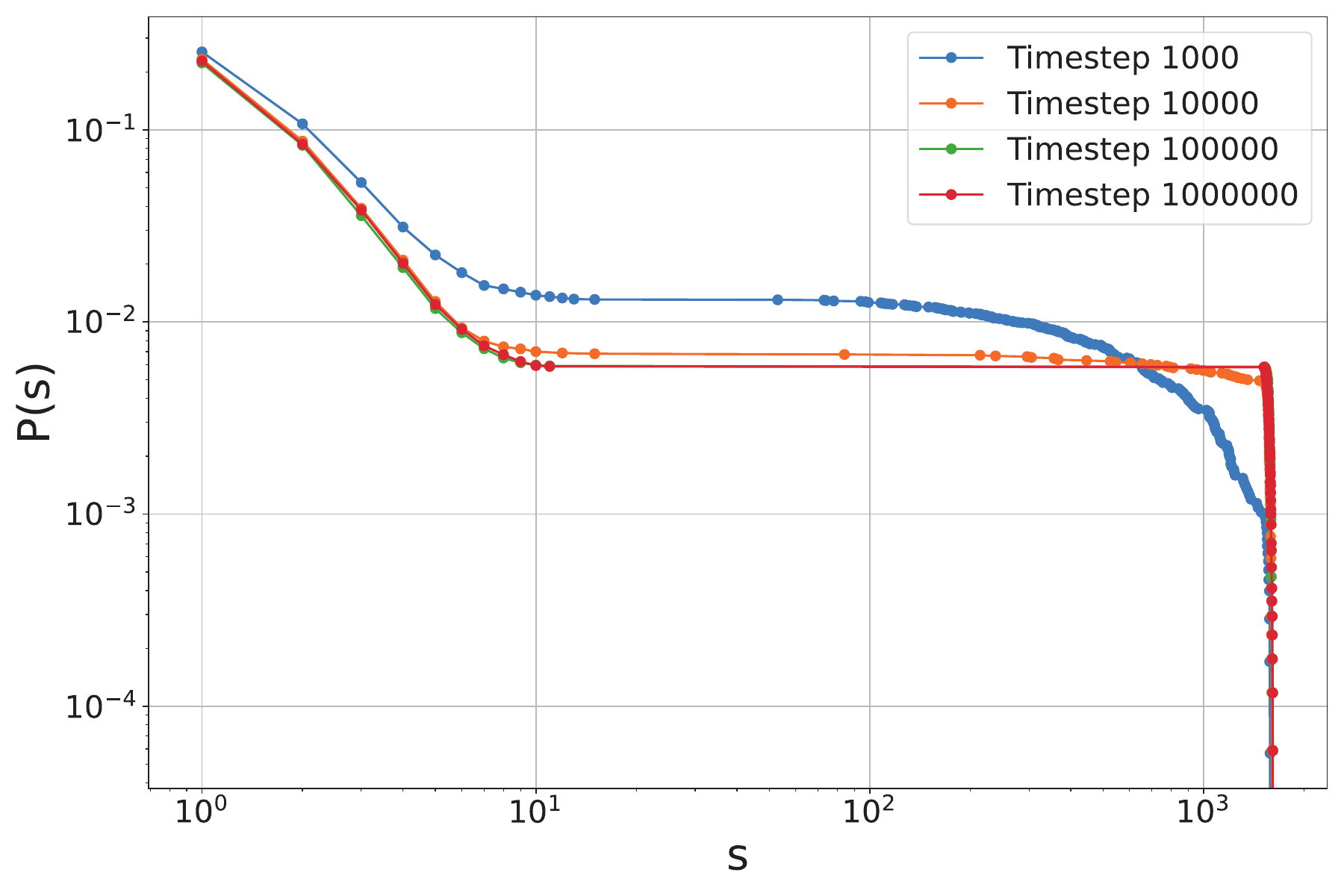}}
\subfigure[$L=120$]{\includegraphics[width=0.47\linewidth]{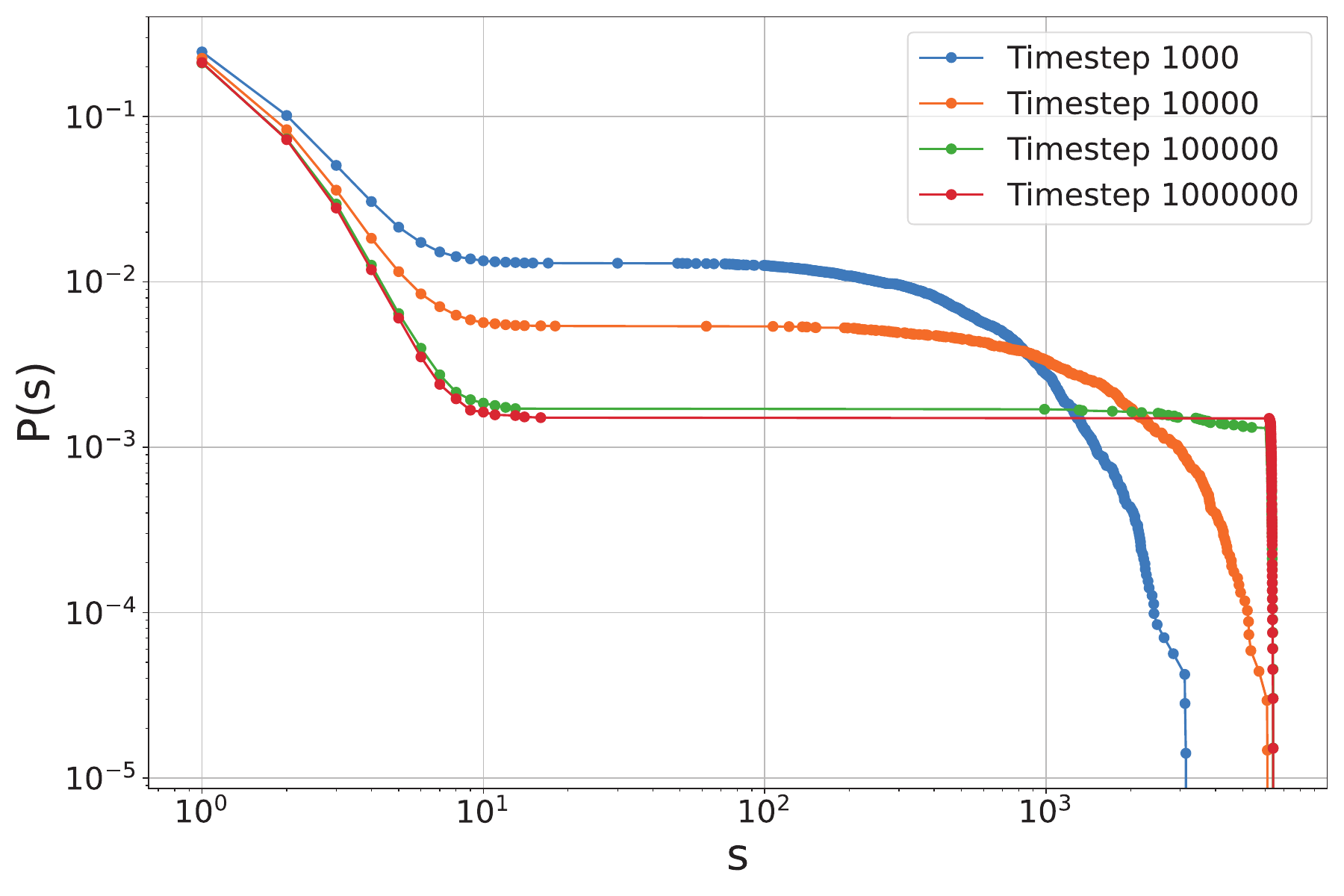}} 
\subfigure[$L=180$]{\includegraphics[width=0.47\linewidth]{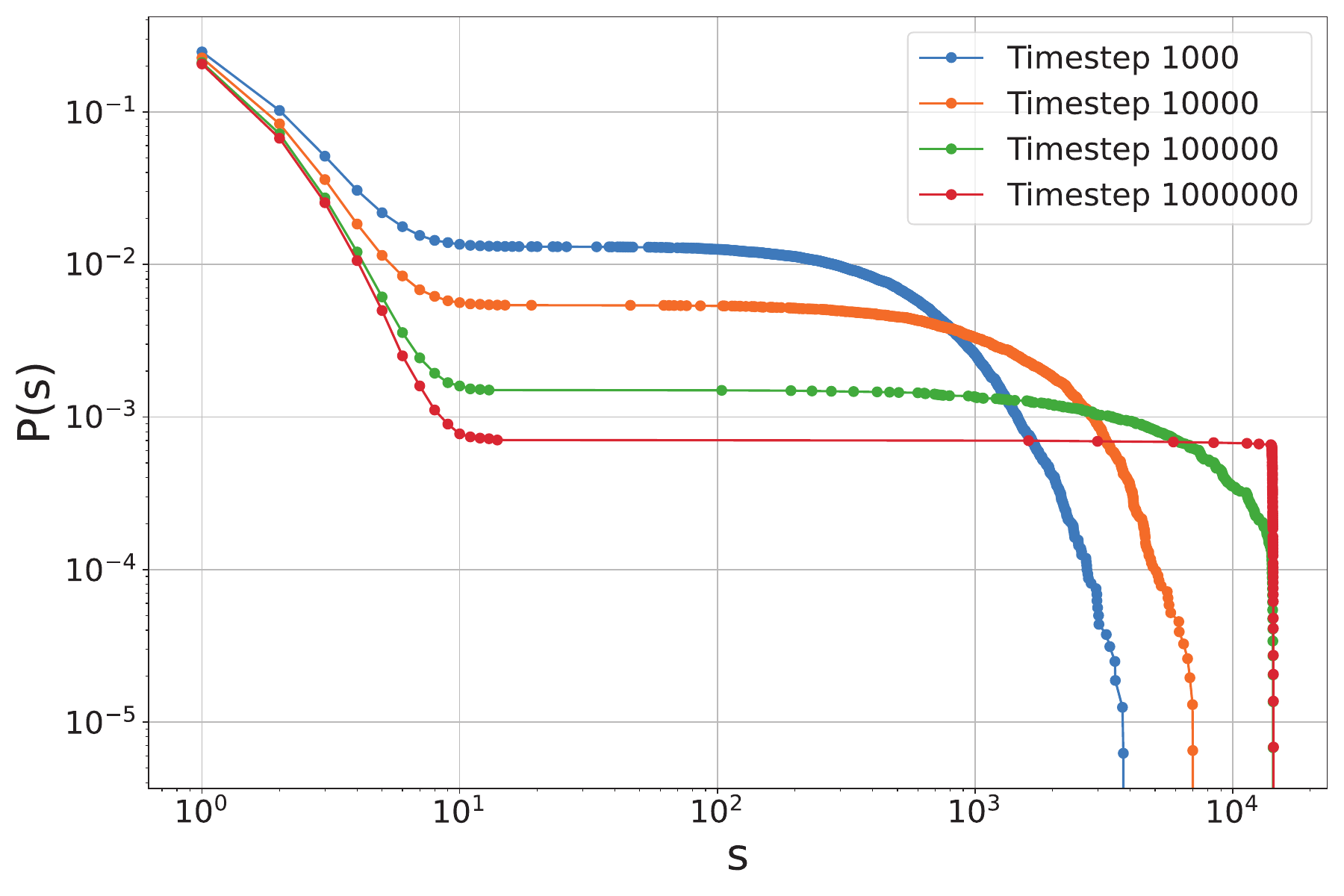}}
\subfigure[$L=240$]{\includegraphics[width=0.47\linewidth]{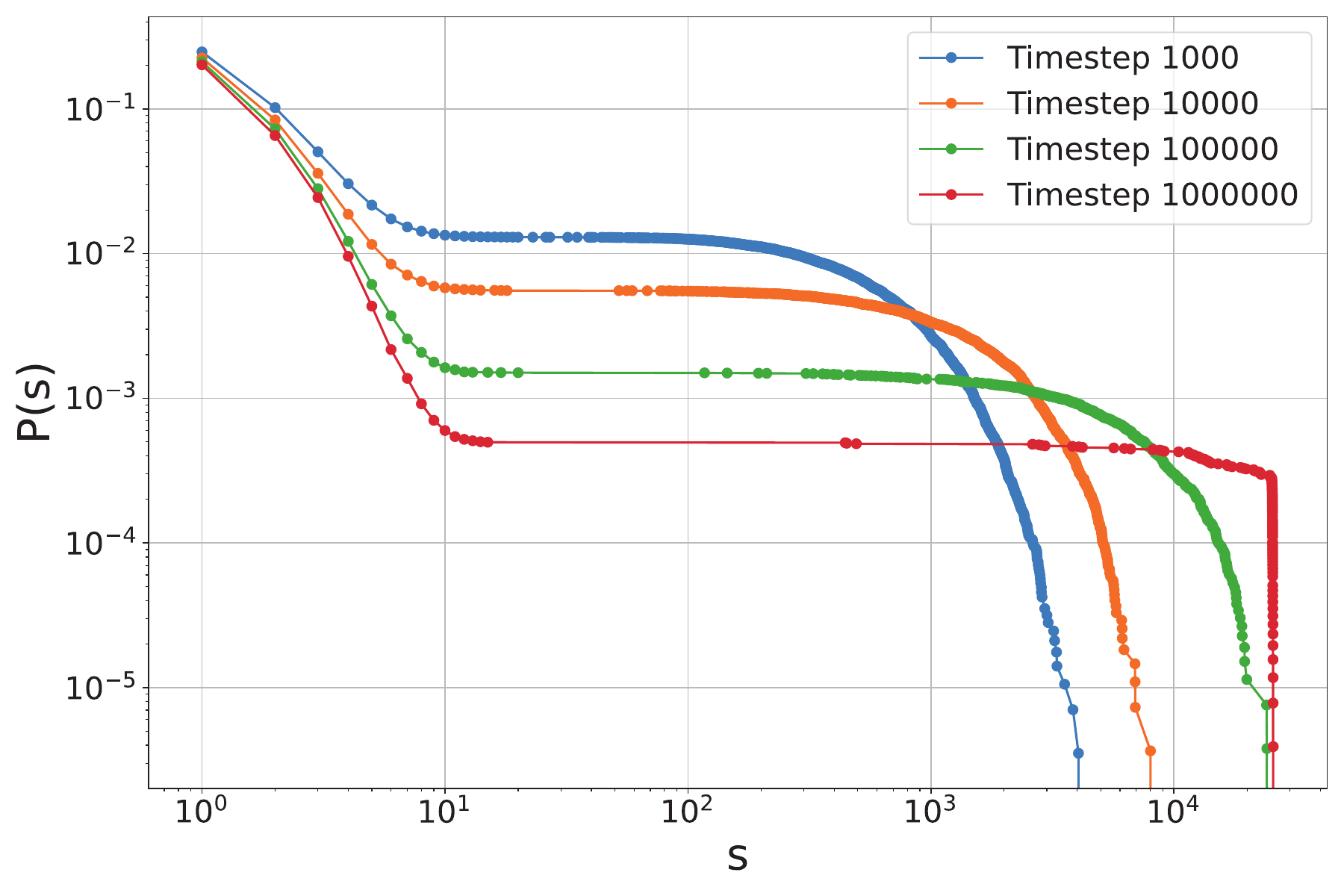}}  
\caption{(Color online) Time evolution of the cumulative distribution of cluster sizes for $\rho=0.5$ for different system sizes $L$. The distributions are obtained by averaging over $100$ samples. The horizontal axis represents the cluster size $s$, and the vertical axis represents the cumulative distribution $P(s)$ defined in the text. The distributions at four different time steps, $t=1\,000$, $10\,000$, $100\,000$, and $1\,000\,000$, are superimposed for comparison: (a) $L=60$, (b) $L=120$, (c) $L=180$, and (d) $L=240$.}
\label{fig:cluster_distribution_L}
\end{figure}
Below, we quantitatively evaluate the cluster size distribution at various times using the Hoshen-Copelman algorithm \cite{hoshen1976percolation} for the cluster identification. In this algorithm, particles assigned the same label are defined to belong to the same cluster, and the cluster size is denoted by $S$ at a given time. Let $M(S)$ be the number of clusters of size $S$ at the given time. The normalized number $p(S)$ is then defined as
\begin{equation}
  p(S) = \dfrac{M(S)}{\displaystyle \int_0^\infty M(S')\,\mathrm{d}S'},
\end{equation}
and the cumulative distribution $P(S)$ is given by
\begin{equation}
  P(S) = \int_S^\infty p(S')\,\mathrm{d}S' 
        = 1 - \int_0^S p(S')\,\mathrm{d}S'.
\end{equation}
Figure~\ref{fig:cluster_distribution_L} shows the cumulative distribution of cluster sizes for the case of particle density $\rho=0.5$, obtained by averaging over $100$ samples for several system sizes.

In all system sizes, at early time steps of around $t=1\,000$, small clusters dominate and large clusters are scarcely observed. However, cluster merging and growth occur as time progresses, and the cumulative distribution shifts toward larger sizes. 
Regardless of the system size $L$, a single giant cluster eventually occupies most of the system after a sufficiently long time. 
Beyond a specific time, the maximum cluster size converges to a distinct value for each system size. 
This behavior indicates an upper bound to the size of the largest cluster formed via motility-induced phase separation in the SPLG-AXY model.

Notably, the maximum cluster size changes with system size is particularly interesting. 
Due to the relatively low number of particles in small systems, the maximum cluster size remains small even after cluster merging.
In contrast, in larger systems, the greater number of particles leads to more extensive merging and growth, resulting in a tendency for the maximum cluster size to be larger. 
This behavior is evident from the cumulative distributions in Fig.~\ref{fig:cluster_distribution_L}: The curve for $L=240$ extends further to the right, while that for $L=60$ reaches the maximum cluster size at an earlier stage with a less pronounced tail.

Furthermore, from the growth rate perspective, larger systems require more MCSs to form a giant cluster. This behavior is consistent with the intuition that as the system size increases, the relaxation time becomes longer due to the process in which small clusters meet, merge, and grow.

\begin{figure}
\centering
\subfigure[$L=60$]{\includegraphics[width=0.47\linewidth]{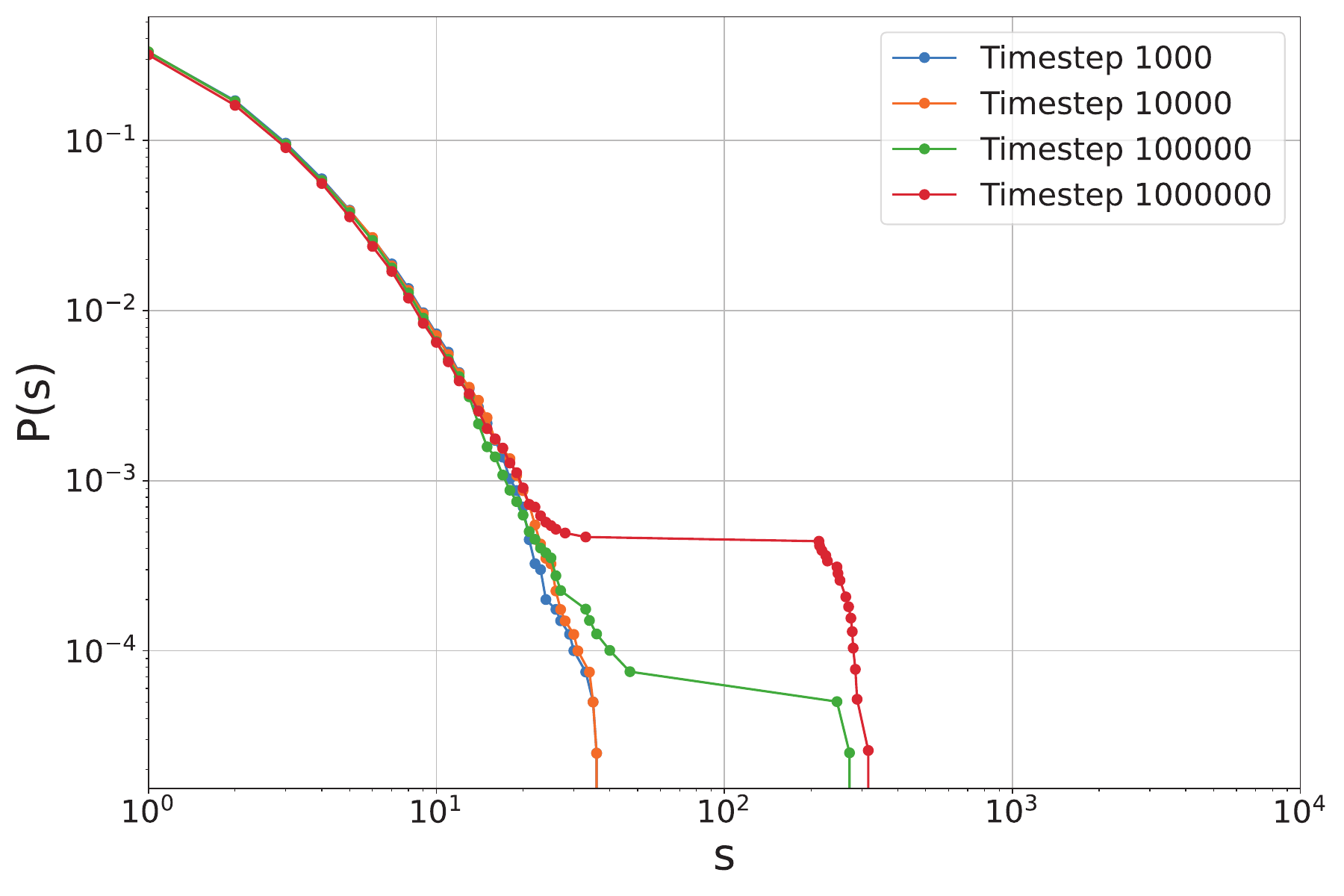}}
\subfigure[$L=120$]{\includegraphics[width=0.47\linewidth]{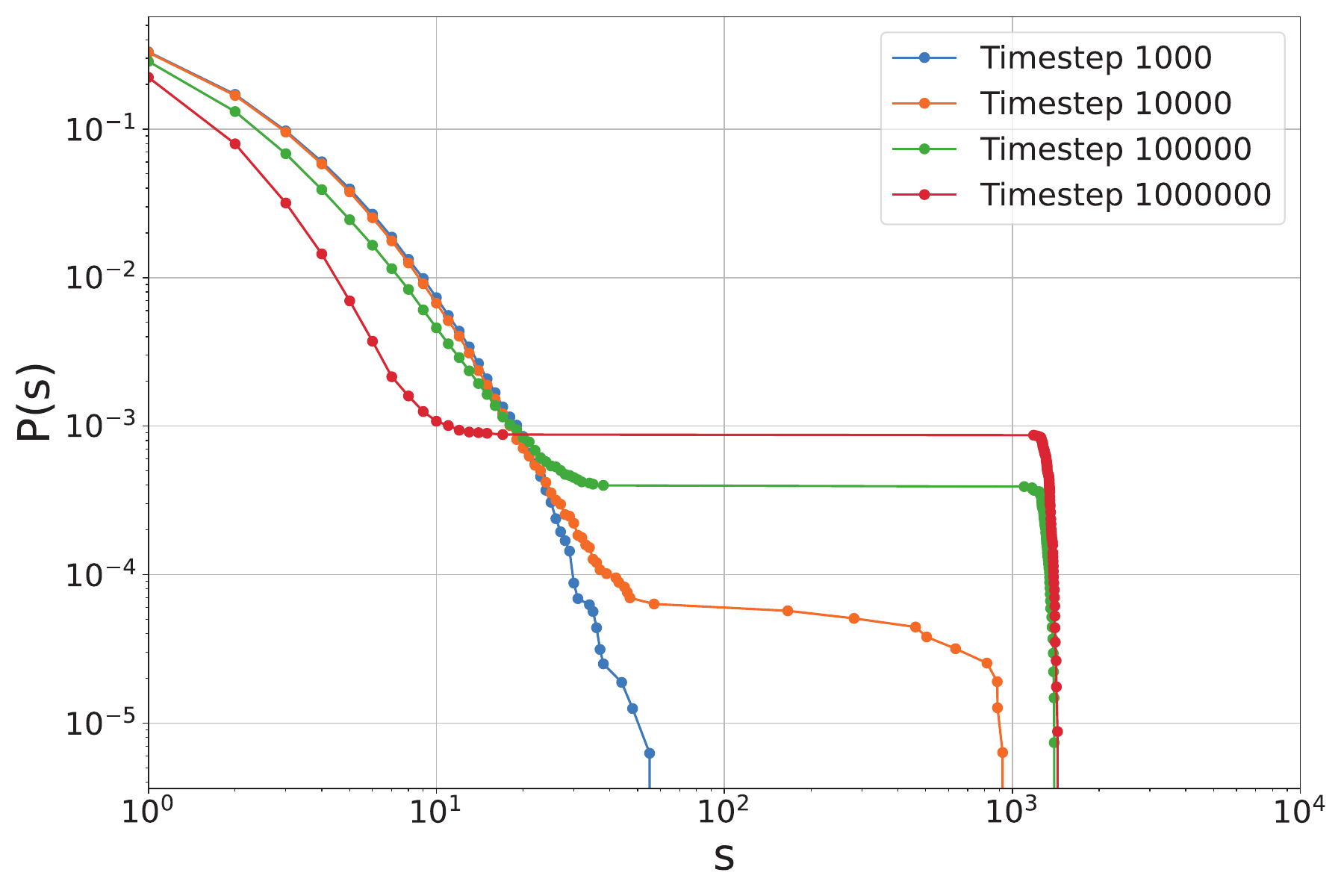}} 
\subfigure[$L=180$]{\includegraphics[width=0.47\linewidth]{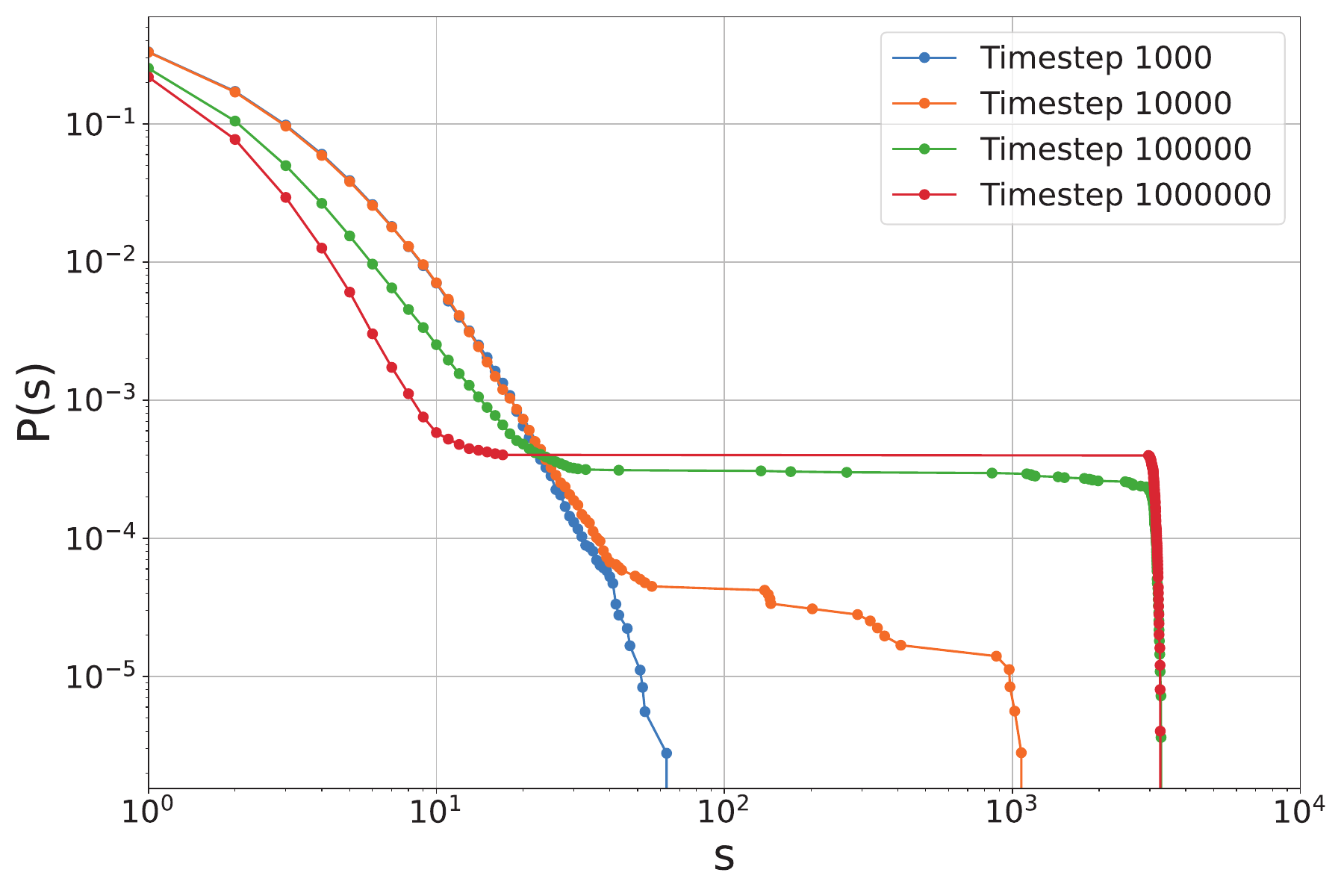}}
\subfigure[$L=240$]{\includegraphics[width=0.47\linewidth]{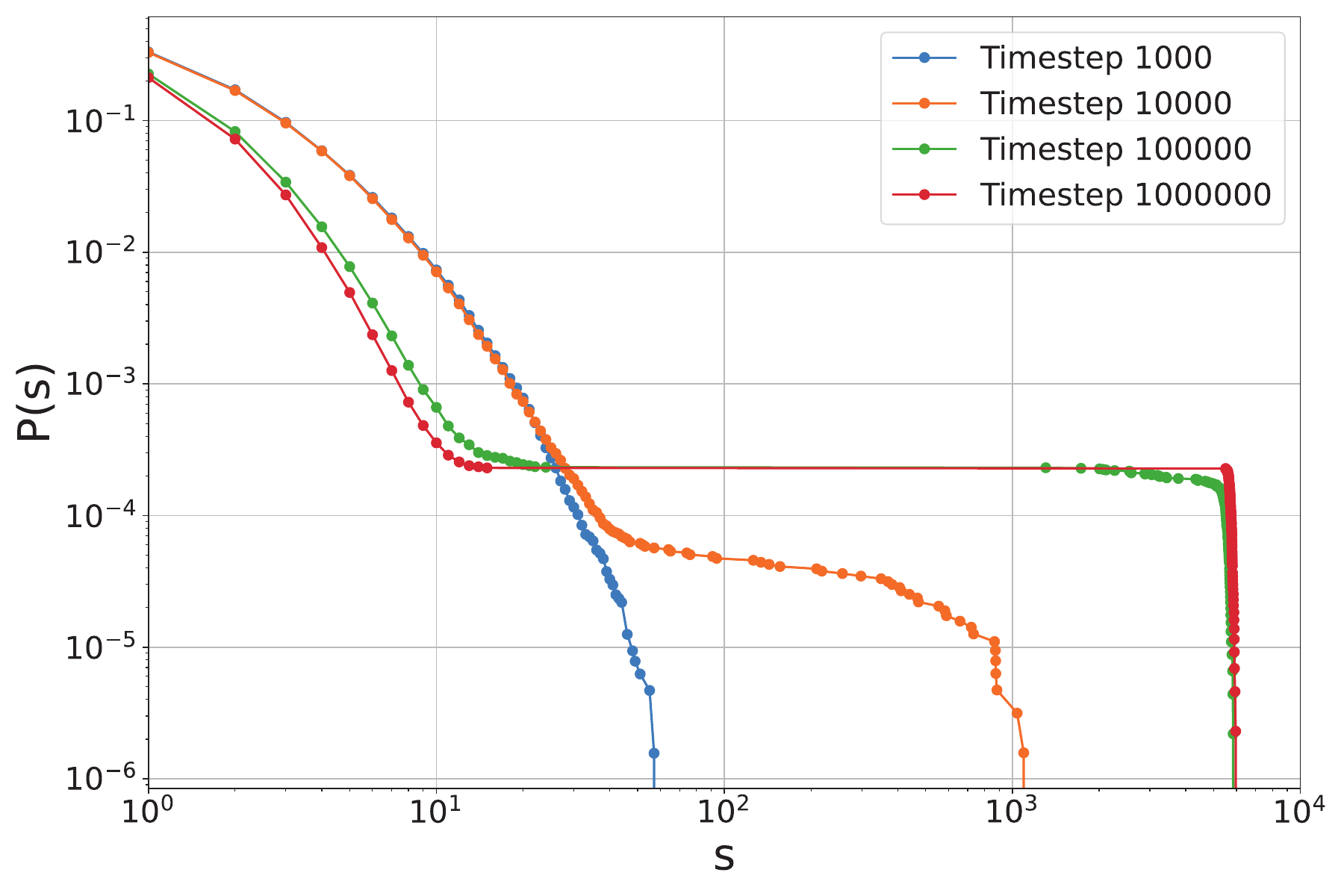}}
\caption{(Color online) Time evolution of the cumulative distribution of cluster sizes for $\rho=0.2$ for different system sizes $L$. The distributions are obtained by averaging over $100$ samples. The horizontal and the vertical axes, time steps, and system sizes are the same as shown in Fig.~\ref{fig:cluster_distribution_L}.}
\label{fig:cluster_distribution_L_rho20}
\end{figure}
Under similar conditions, the cumulative distribution of the cluster sizes for the case of low particle density $\rho=0.2$, obtained by averaging over $100$ samples, 
is shown in Fig.~\ref{fig:cluster_distribution_L_rho20}.
For the low-density case $\rho=0.2$, it is found that when the system size is too small, for instance, $L=60$, the maximum cluster size does not stabilize sufficiently.
After sufficient time, the cumulative distribution typically converges to a shape with a sharp drop at the maximum cluster size. 
However, as observed in Fig.~\ref{fig:cluster_distribution_L_rho20}(a), for $\rho=0.2$ and $L=60$, even after a long time, the distribution does not exhibit a pronounced concentration corresponding to the maximum cluster size.
This behavior indicates a minimal size of the possible or stable cluster independent of the system size, and that the maximum cluster size for $\rho=0.2$ with $L=60$ is close to this minimal size, rendering it unstable.

\subsubsection{Density Dependence of the Maximal Cluster Size}
\label{subsec:cluster_function}

In Sec.~\ref{subsec:cluster_distribution}, we show that the maximum cluster size in the steady state becomes large as the system size $L$ increases. However, since the total number of particles also increases with the system size at a fixed density, we must investigate whether the occupancy fraction of the maximal cluster relative to the entire system depends on the system size $L$.

To this end, we define the area fraction of the maximum cluster size as
\begin{equation}
  \phi_{\mathrm{max}} 
  = \frac{\langle S_\mathrm{m} \rangle}{L^2}
  \label{eq:max_cluster_ratio}
\end{equation}
where $\langle S_{\mathrm{m}} \rangle$ is the average maximum cluster size measured in the steady state, and $L^2$ denotes the total area of the system. The parameter $\phi_{\mathrm{max}}$ indicates the fraction of the system's area occupied by the largest cluster, and it is a crucial metric for investigating the scaling behavior as the system size $L$ is varied.

\begin{figure}
\centering
\includegraphics[width=1.0\linewidth]{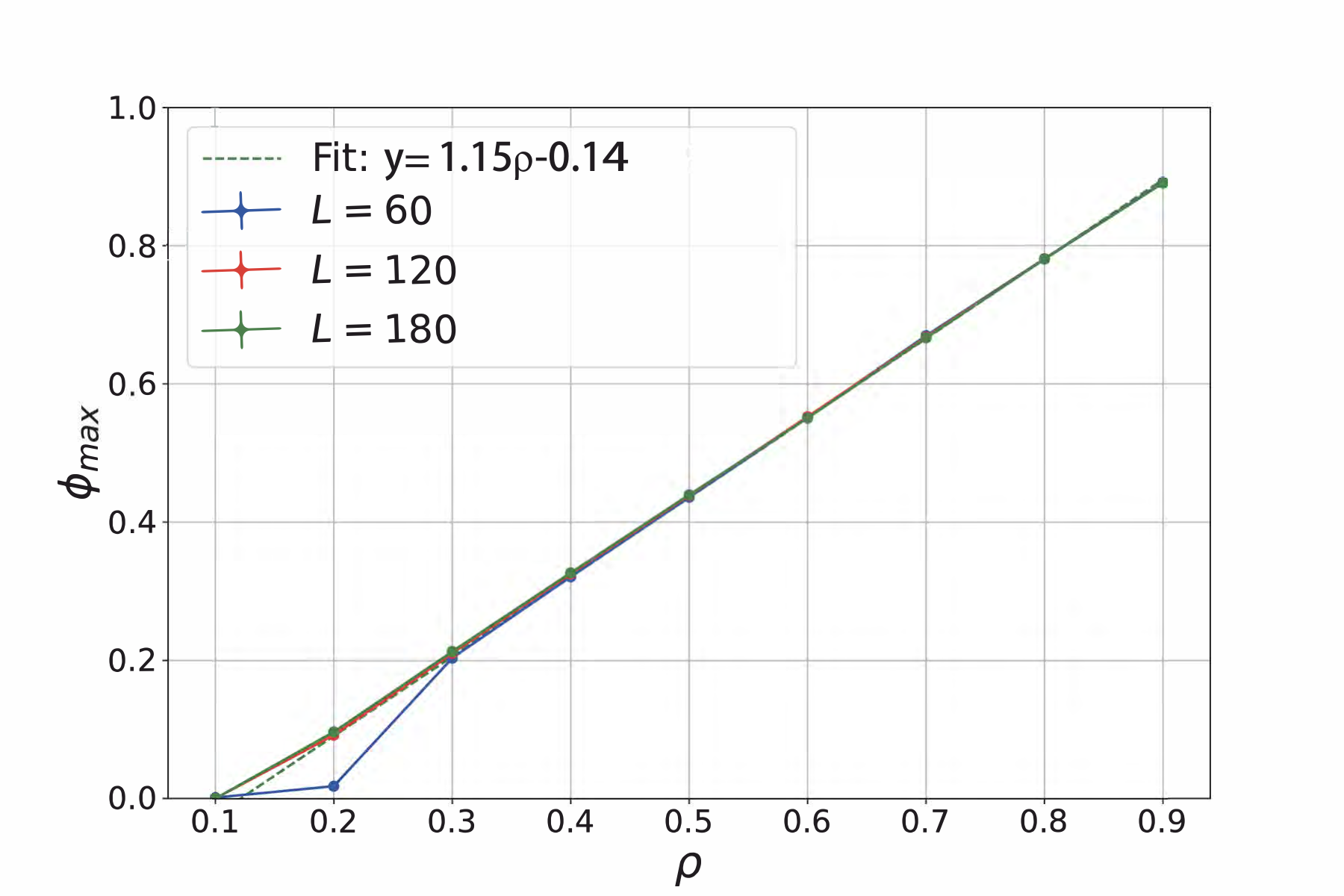}
\caption{(Color online) Plot of the area fraction of the maximum cluster size, $\phi_{\mathrm{max}}=\langle S_{\mathrm{m}} \rangle / L^2$, as a function of the particle density $\rho$. The cases for system sizes $L=60$, $120$, and $180$ are compared. The result of the linear fitting, $\phi_{\mathrm{max}} \simeq 1.15 \rho - 0.14$, is also shown. $\langle S_{\mathrm{m}} \rangle$ is averaged over $100$ samples after the relaxation of $2\,000\,000$ MCSs. The error bar of each data point is smaller than the size of the markers.}
\label{fig:max_cluster_fraction}
\end{figure}
Figure~\ref{fig:max_cluster_fraction} shows the area fraction of the maximum cluster size defined by Eq.~\ref{eq:max_cluster_ratio} for various particle densities $\rho$, and several system sizes, $L=60$, $120$, and $180$. The area fraction of the maximum cluster size is calculated after relaxation of $2\,000\,000$ MCSs, and it is averaged over $100$ samples.

From Fig.~\ref{fig:max_cluster_fraction}, it can be seen that the value of $\phi_{\mathrm{max}}$ does not depend on the system size $L$ except for small densities. This behavior indicates that the fraction of the system's area occupied by the maximum cluster in the steady state remains nearly constant regardless of $L$. In general, as the system size increases, the number of particles increases and the maximum cluster size $\langle S_{\mathrm{m}} \rangle$ becomes large; However, its ratio relative to the total area $L^2$ does not increase, remains constant. 
This result represents one of the essential characteristics of the SPLG-AXY model.

Furthermore, when we see the area fraction of the maximum cluster size as a funciton of the particle density $\rho$,  $\phi_{\mathrm{max}}$ is found to be in good agreement with a linear fit, namely $\phi_{\mathrm{max}} \simeq 1.15\,\rho - 0.14$. 
This behavior indicates that the particle density primarily determines the area fraction of the maximum cluster. 
There is a slight deviation from the linear function in the low-density region, particularly for the small system size. This deviation can be explained by the fact that the area fraction of the maximum cluster is close to the minimum cluster size that can exist in this region.

These results resemble the gas-liquid coexistence phase of the first-order phase transition. 	
We denote the total number of particles as $n$, the number of particles belonging to the maximum cluster as $n_{m}$, 
and the number of the remaining particles, $n_{g}$, which are isolated from the maximum cluster and randomly move around it. 
Then the conservation of the particles gives the following relation:
\begin{equation}
n_{m} + n_{g} = n.
\end{equation}
Dividing this relation by the total area $L^{2}$ and inserting the individual area of the region, $A_{m}$ and $A_{g}$, that is, the aggregated cluster region and random walking region, respectively, we obtain the following relation:
\begin{equation}
\dfrac{n_{m}}{A_{m}}\dfrac{A_{m}}{L^{2}} + \dfrac{n_{g}}{A_{g}}\dfrac{A_{g}}{L^{2}}=\rho.
\end{equation}
Because the density of the maximum cluster $\rho_{m} = n_{m}/A_{m}$ is $1$, and we denote the particle density of the random walking region as $\rho_{g} = n_{g}/A_{g}$, we obtain the linear relation between $\phi_{\mathrm{max}}$ and $\rho$ as
\begin{equation}
\phi_{\mathrm{max}} = \dfrac{1}{1-\rho_{g}} \left( \rho - \rho_{g} \right), 
\end{equation}
where we use $A_{m}/L^{2} = \phi_{\mathrm{max}}$ and $A_{g}/L^{2} = 1-\phi_{\mathrm{max}}$, which are the definition of $\phi_{\mathrm{max}}$.
The fitting result gives $\rho_{g} \simeq 0.13$. 
This density represents the critical density below which cluster aggregation does not occur at the thermodynamic limit. 
From the deviation from the linear relation at $\rho=0.2$ of $L=60$ and the coincidence for $\rho=0.3$ of $L=60$ in Fig.~\ref{fig:max_cluster_fraction}, the minimal stable cluster size $n_{m}^{(s)}$ is approximately estimated as $ 300 < n_m^{(s)} < 700$.

\subsection{Time Evolution of the Maximum Cluster Size}

\begin{figure}
\centering
\subfigure[$\rho = 0.3$]{\includegraphics[width=0.48\linewidth]{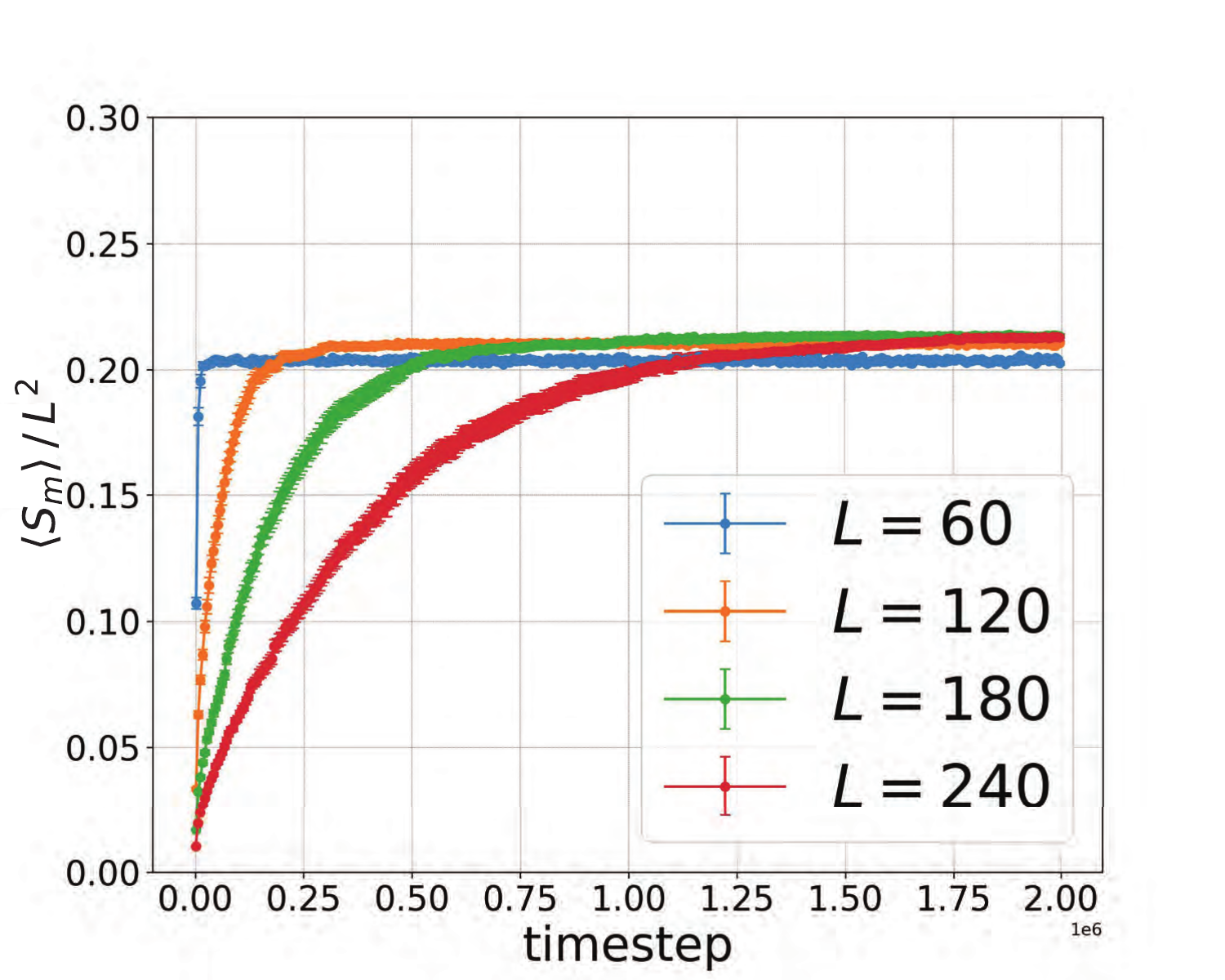 }}
\subfigure[$\rho = 0.4$]{\includegraphics[width=0.48\linewidth]{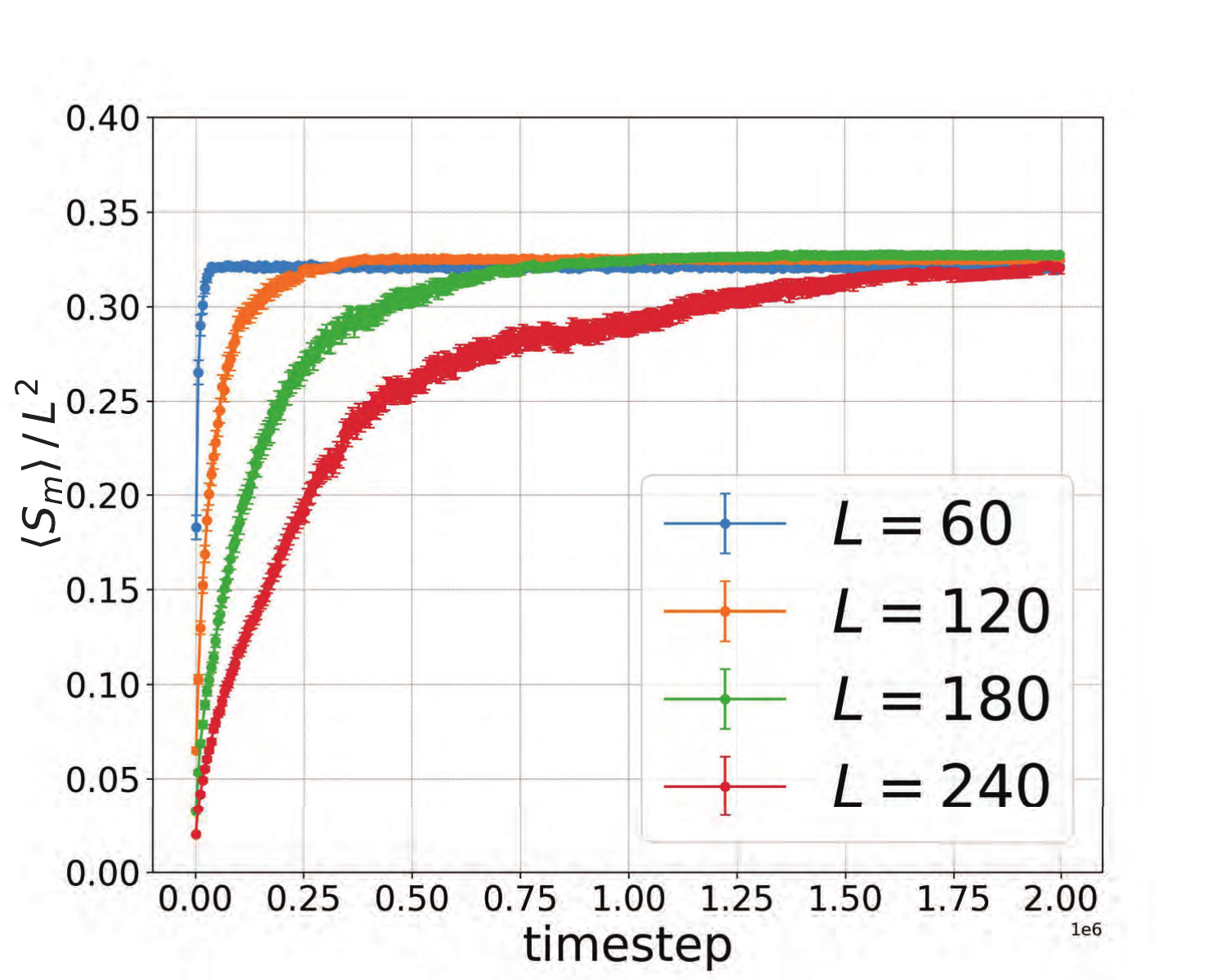 }} 
\subfigure[$\rho = 0.5$]{\includegraphics[width=0.48\linewidth]{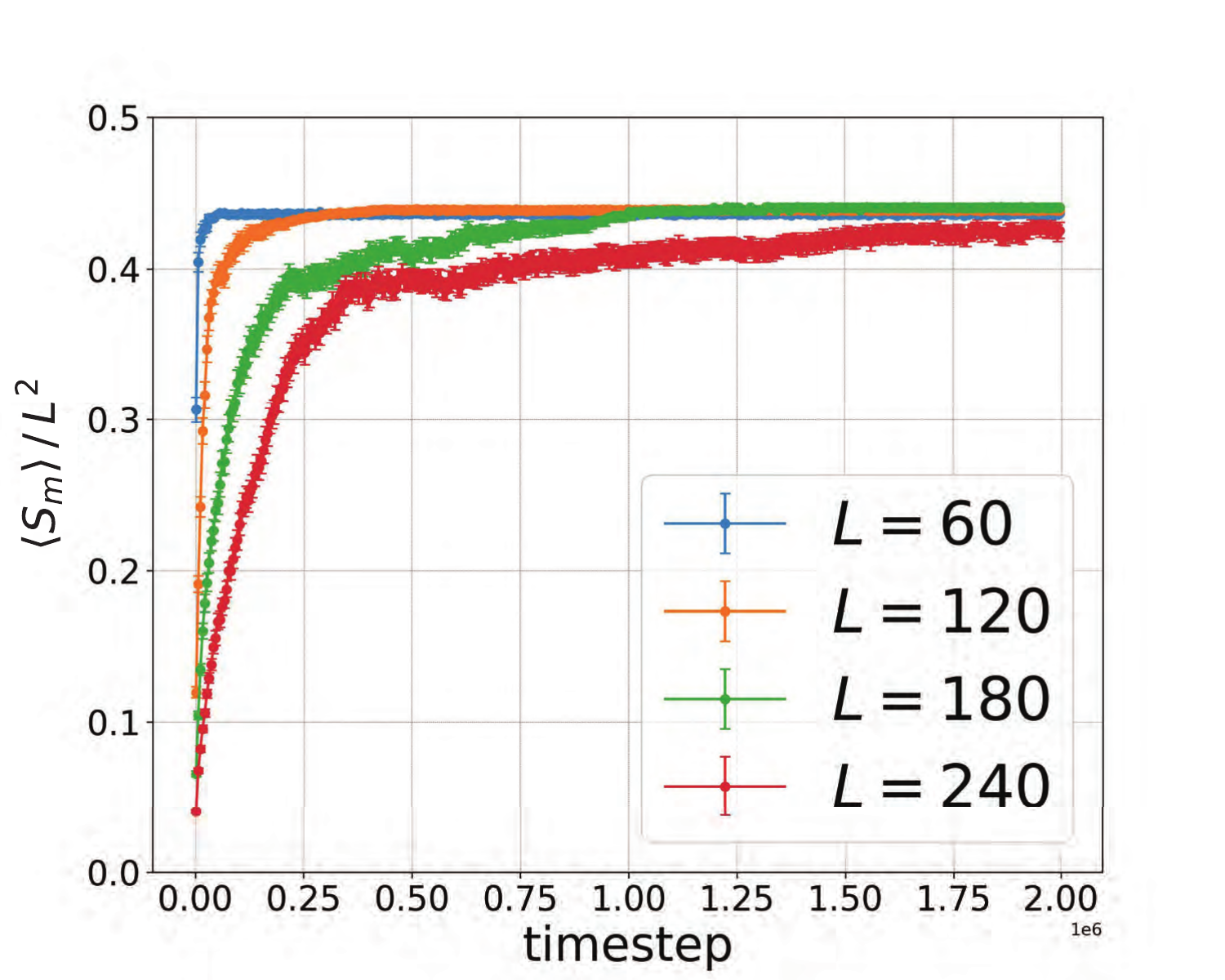 }}
\subfigure[$\rho = 0.6$]{\includegraphics[width=0.48\linewidth]{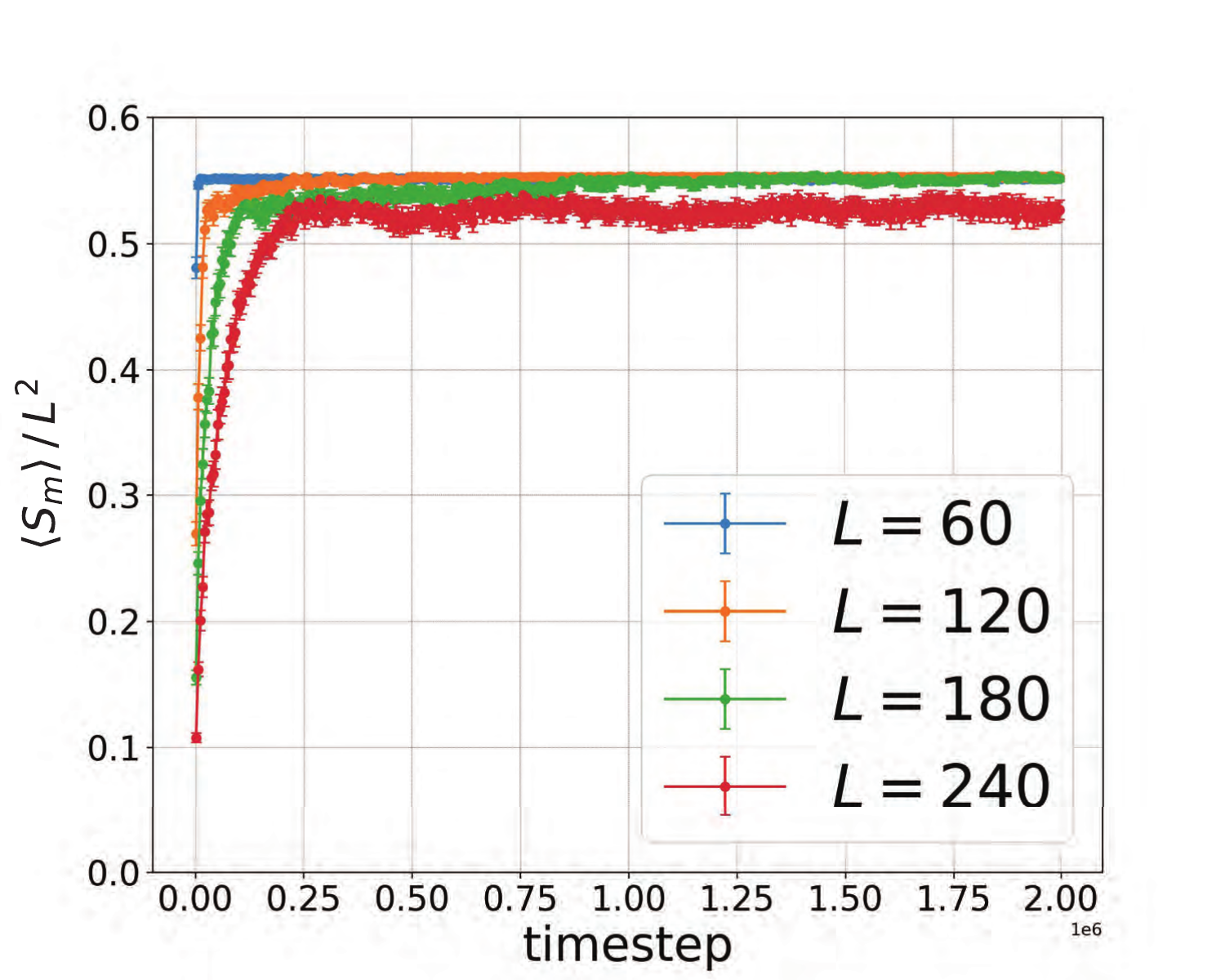 }}
\caption{(Color online) Time evolution of the maximum cluster size. The horizontal axis represents the number of MCSs in units of $10^{6}$ MCSs, and the vertical axis represents the area fraction of the maximum cluster, $\phi_{\mathrm{max}} = \langle S_{\mathrm{m}} \rangle / L^2$. The data for system sizes $L=60, 120, 180,$ and $240$ are compared for several densities: (a) $\rho = 0.3$, (b) $\rho = 0.4$, (c) $\rho = 0.5$, and (d) $\rho = 0.6$.}
\label{fig:cluster_timestep_L}
\end{figure}
We now consider the time evolution of the area fraction of the maximum cluster size $\phi_{\mathrm{max}}$. Given that the area fraction of the maximum cluster exhibits a universal behavior across system sizes, we expect that its time evolution also contains universal features. To track this evolution, we employ $\phi_{\mathrm{max}} = \langle S_\mathrm{m} \rangle / L^2$, as defined in Section~\ref{subsec:cluster_function}. Figures~\ref{fig:cluster_timestep_L} compare the time evolution of the maximum cluster size for particle densities $\rho = 0.3, 0.4, 0.5,$ and $0.6$ at various system sizes. It is noted that at lower densities, a single maximum cluster does not form. The algorithm faces difficulties distinguishing clusters at higher densities due to randomly aggregated particles and vortex defects. Therefore, we focus on the above four densities. The simulations were performed for $2\,000\,000$ MCSs, and the data are averaged over $100$ samples.

From Fig.~\ref{fig:cluster_timestep_L}, we get the following observations. First, as the system size $L$ increases, the relaxation time required to reach the final steady state (in which the maximum cluster size becomes nearly constant) increases. On the other hand, the final value of $\phi_{\mathrm{max}}$ converges to nearly the same value for a given particle density $\rho$, as discussed in the previous section. This result is consistent with the intuitive interpretation that while larger systems require more time to aggregate a single large cluster from the initial uniform distribution, the final occupancy fraction in the steady state is determined solely by the density of the particles. Moreover, as the density increases, the initial rapid increase in cluster growth becomes more pronounced.

Here, we perform a scaling transformation on the timesteps of Fig.~\ref{fig:cluster_timestep_L} by dividing them by $L^3$. This scaling assumes that the relaxation time $\tau$ is governed by a scaling law of the form $\tau \sim L^3$. This ansatz originates from the scaling behavior of the conserved order parameter in the coexistence phase of a first-order phase transition, known as the Lifshitz-Slyozov law \cite{LS1961a, CL1995}; A spatial characteristic scale $\xi$ of the conserved order parameter cluster exhibits power-law growth with time $t$ as $\xi \sim t^{1/3}$. 

\begin{figure}
\centering
\subfigure[$\rho = 0.3$]{\includegraphics[width=0.48\linewidth]{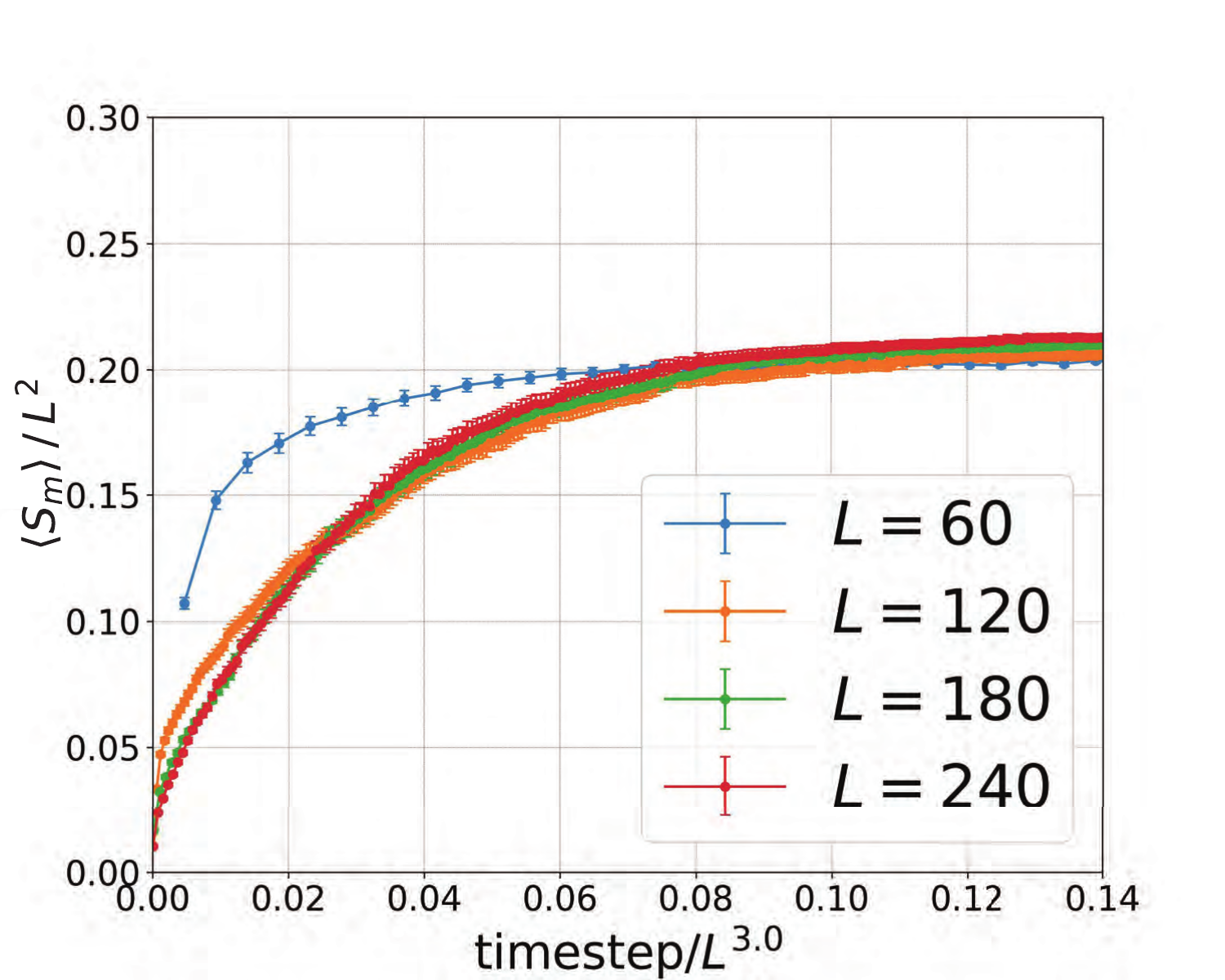}}
\subfigure[$\rho = 0.4$]{\includegraphics[width=0.48\linewidth]{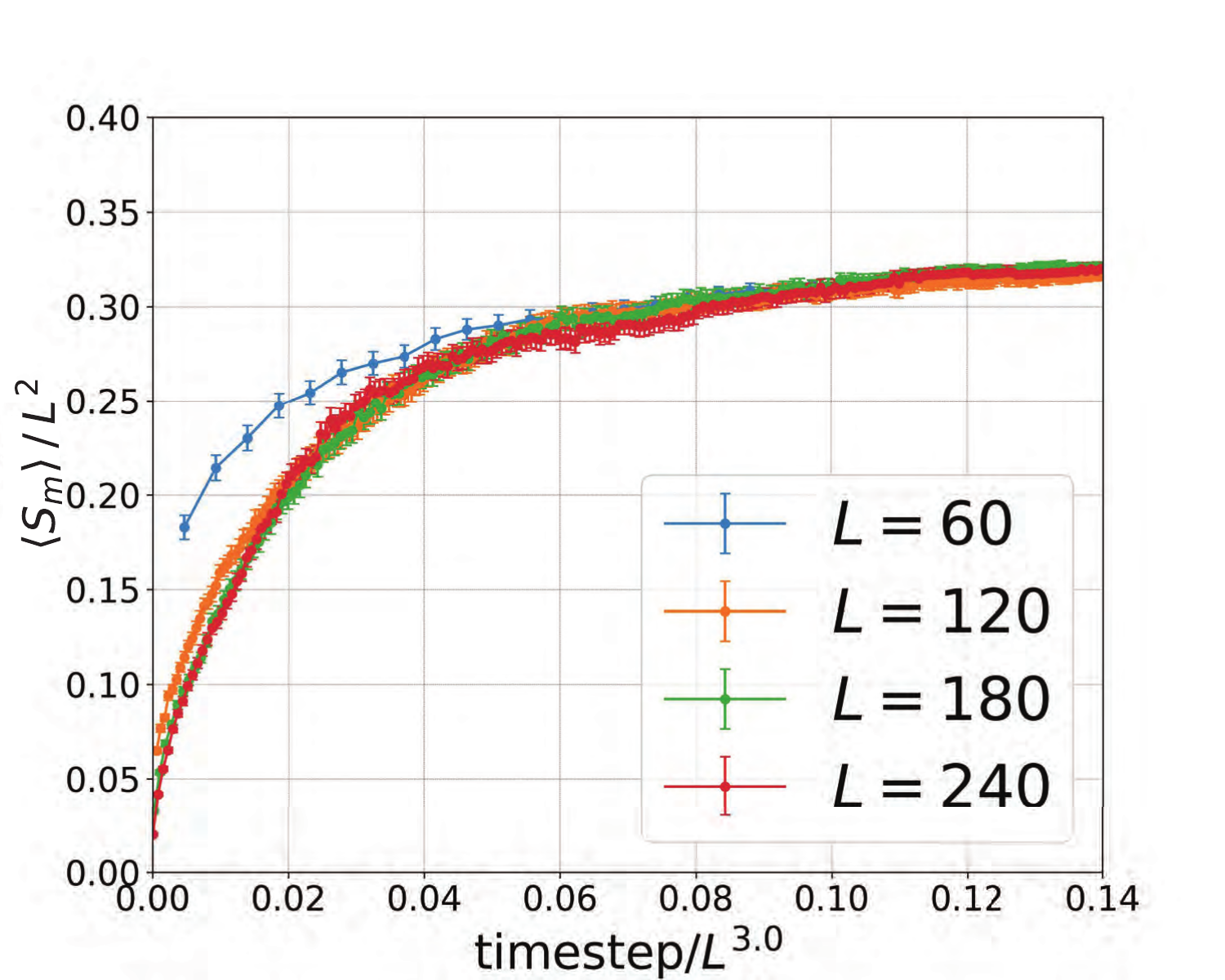}}
\subfigure[$\rho = 0.5$]{\includegraphics[width=0.48\linewidth]{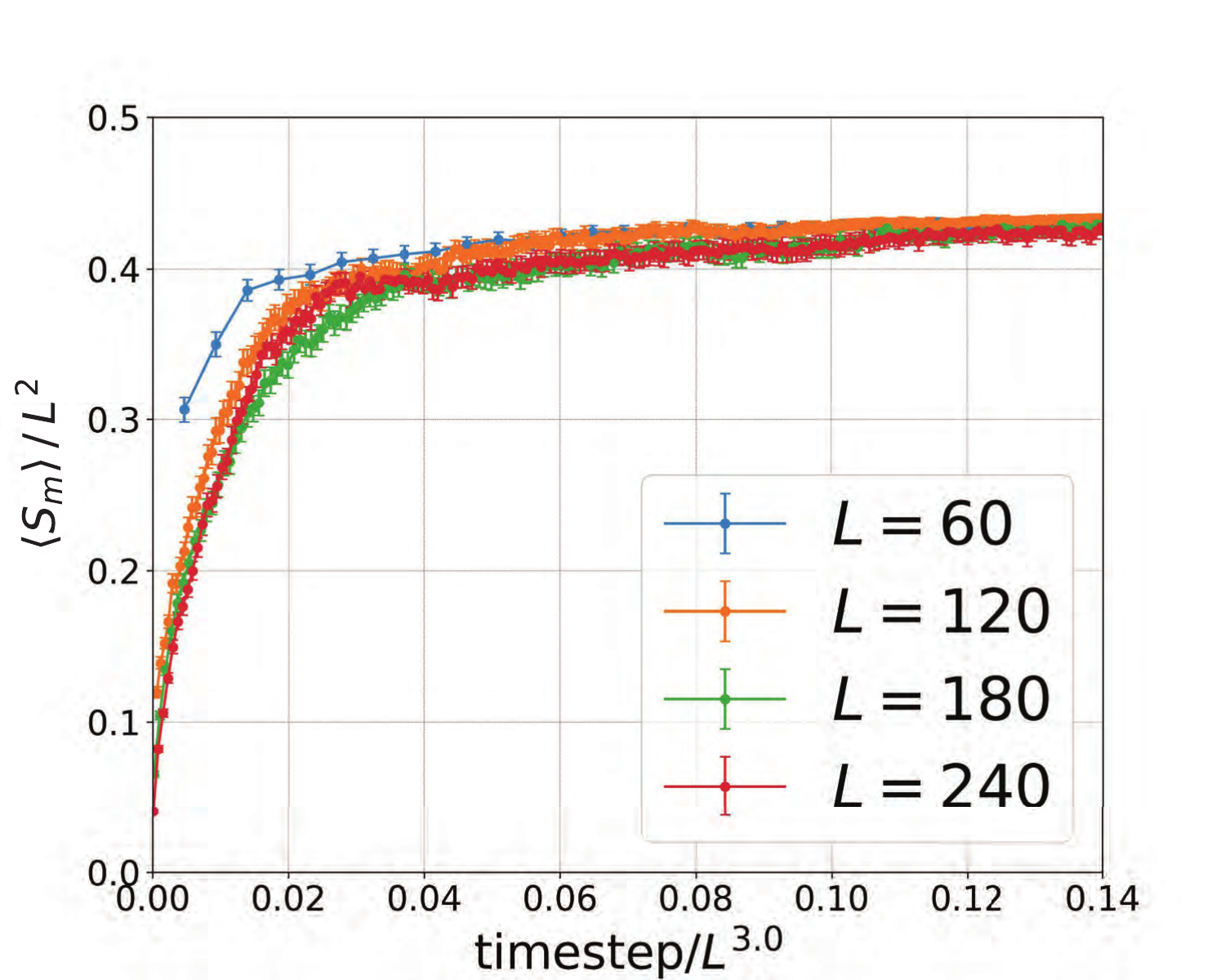}}
\subfigure[$\rho = 0.6$]{\includegraphics[width=0.48\linewidth]{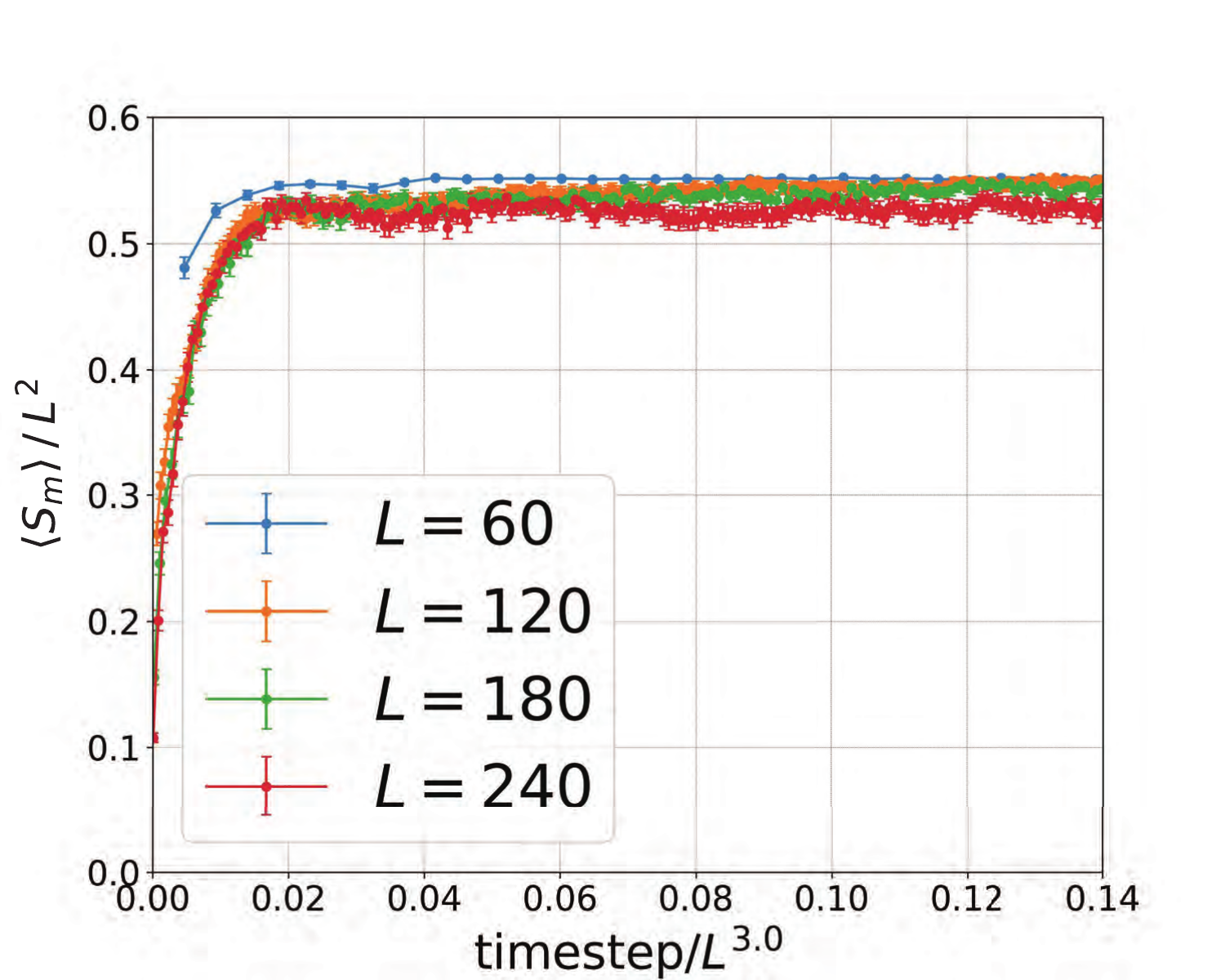}}
\caption{(Color online) Scaling transformation of Fig.~\ref{fig:cluster_timestep_L}. The horizontal axis is redefined as $\text{(MCSs)}/L^3$, and the vertical axis shows the area fraction of the maximum cluster size, $\phi_{\mathrm{max}} = \langle S_{\mathrm{m}} \rangle / L^2$. The data for system sizes $L=60$, $120$, $180$, and $240$ are compared for several densities: (a) $\rho = 0.3$, (b) $\rho = 0.4$, (c) $\rho = 0.5$, and (d) $\rho = 0.6$.}
\label{fig:cluster_timestep_scale}
\end{figure}
The results in Fig.~\ref{fig:cluster_timestep_scale} indicate that by normalizing the time axis by $L^3$, the curves for different $L$ values tend to collapse onto one single curve (or approximately follow a common scaling law) except for the small-sized system. 

In the equilibrium situation, the spatial characteristic length scale is measured by the peak of the structural factor in the wave-number space. In this case, the length scale $L$ does not coincide with this scale. However, the single maximum cluster extending the system size makes the length scale $L$ and the characteristic length of the same order. Incorporating the previously demonstrated linear dependence of the area fraction of the maximum cluster on density, this result suggests that the present SPLG-AXY model exhibits behavior similar to the phenomena observed in the first-order transition in equilibrium systems.

Finally, we evaluate the functional form of the time evolution of the maximum cluster size. 
In conclusion, cluster growth in the SPLG-AXY model can be described as the sum of two exponential relaxation processes. Each relaxation process characterizes a distinct stage in the growth of clusters. Here, we explain the case of $\rho=0.5$ as a representative density.

\begin{figure}
\centering
\subfigure[Log-log plot]{
\includegraphics[width=0.48\linewidth]{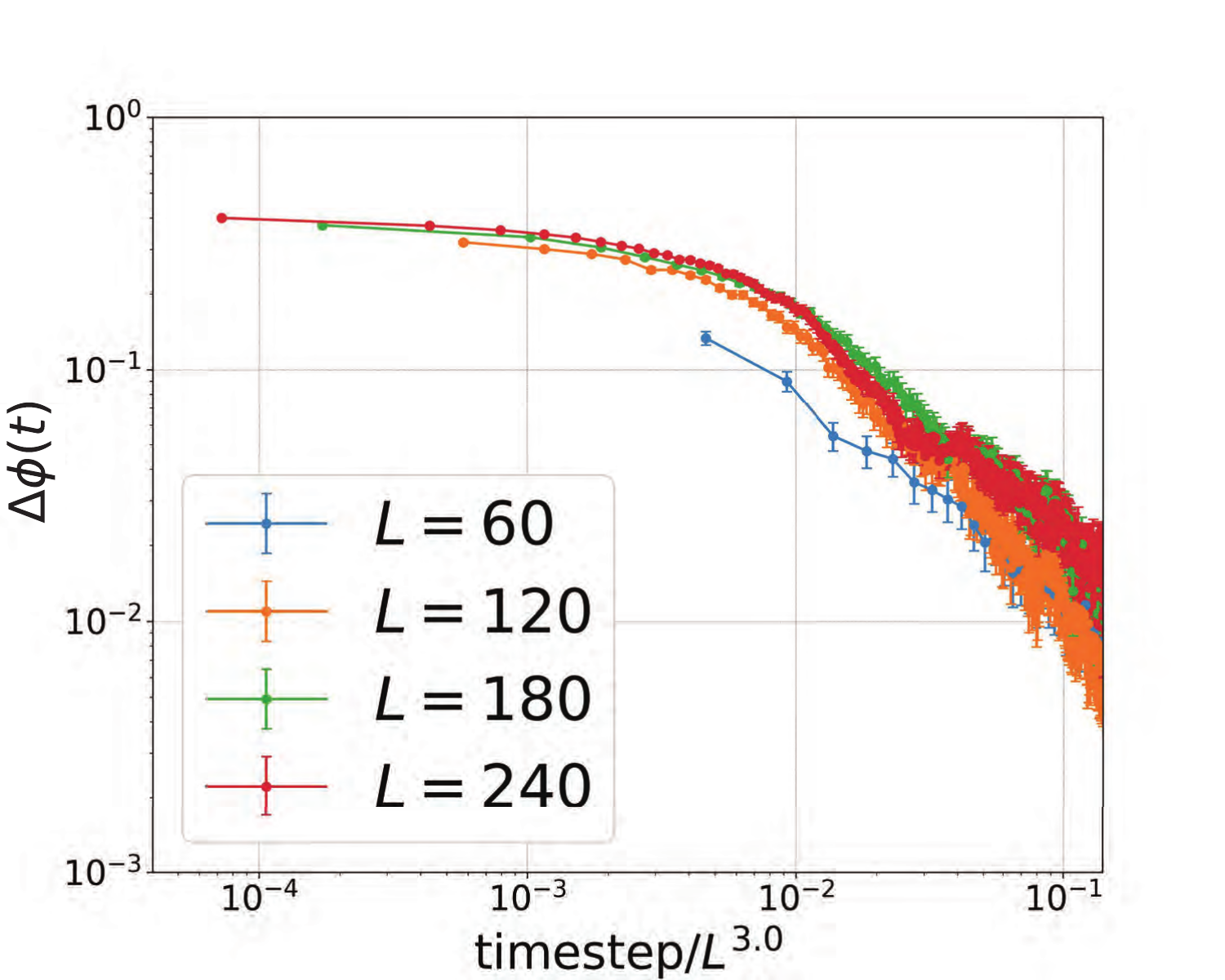}}
\subfigure[Semi-log plot]{
\includegraphics[width=0.48\linewidth]{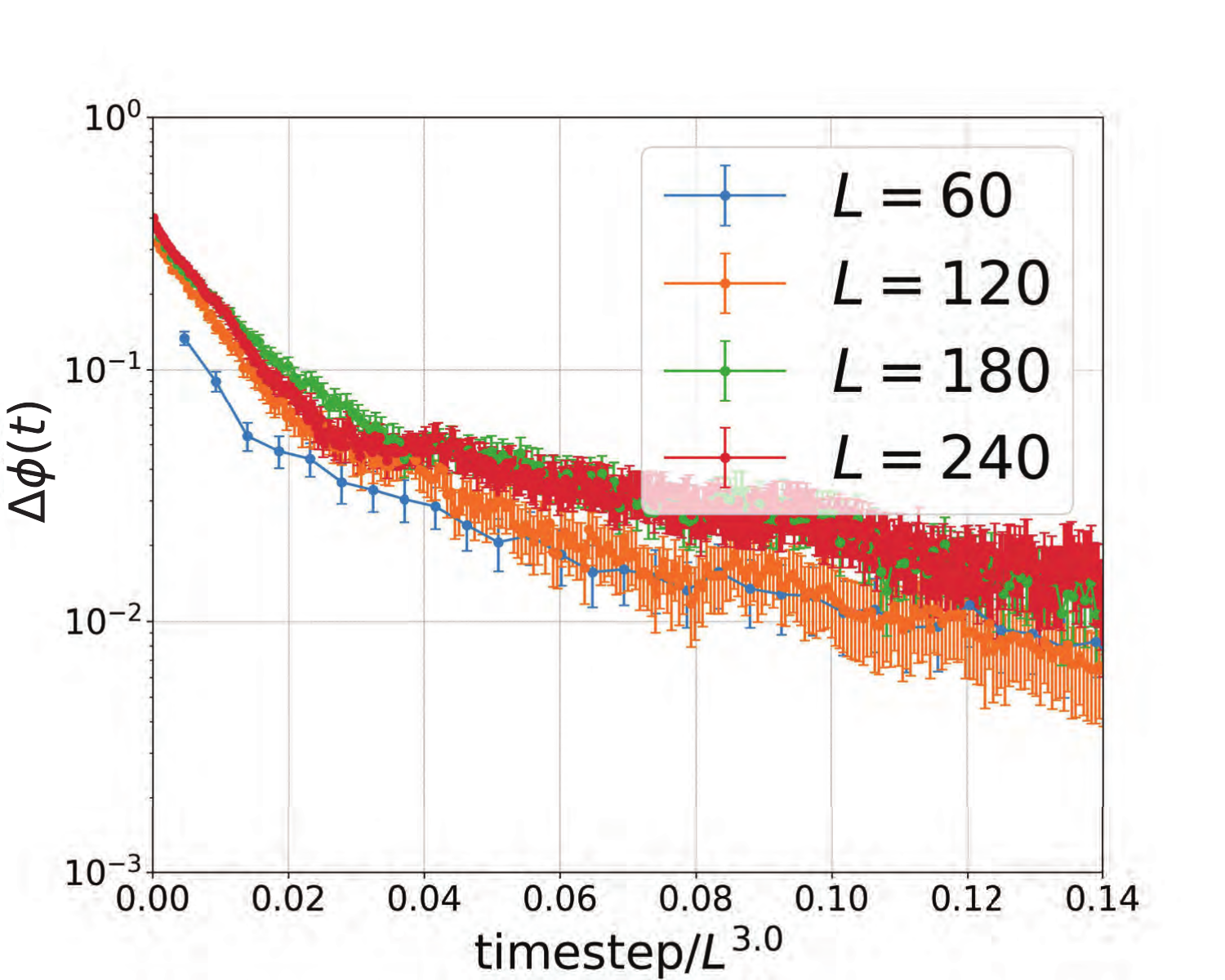}}    
\caption{(Color online) Time evolution of the area fraction of the maximum cluster size $\phi_{\mathrm{max}}(t)$ for $\rho=0.5$, with the time axis scaled by $L^3$. (a) In the double-logarithmic plot, the data would form a straight line if a power-law relaxation is valid, but the results do not support a simple power-law. (b) In the semi-logarithmic plot, two distinct linear regions are observed, corresponding to a rapid initial relaxation and a slower subsequent relaxation, which suggests an exponential decay.}
\label{fig:maxcluster_timeevo_L3}
\end{figure}
Let $\phi_{\mathrm{max}}(t)$ denote the area fraction of the maximum cluster size at time $t$, and let $\phi_{\mathrm{max}}(\infty)$ represent its steady-state value after a sufficiently long time. 
The value $\phi_{\mathrm{max}}(\infty)$ can be obtained by $\phi_{\mathrm{max}} \approx 1.15\,\rho - 0.14$ by the above result.
First, we define
\begin{equation}
\Delta \phi(t) = \phi_{\mathrm{max}}(\infty) - \phi_{\mathrm{max}}(t),
\label{eq:diff}
\end{equation}
which asymptotically approaches zero as $t \to \infty$. 
We can estimate its functional form by plotting $\Delta \phi(t)$ on a double-logarithmic scale and a semi-logarithmic scale.
In the double-logarithmic plot in Fig.~\ref{fig:maxcluster_timeevo_L3}(a), the data would lie on a straight line if a power-law relaxation is valid. However, the results do not exhibit a straight line but rather show curvature as they approach the saturation value. In contrast, in the semi-logarithmic plot shown in  Fig.~\ref{fig:maxcluster_timeevo_L3}(b), two approximately linear regions correspond to the early and later relaxation stages, suggesting an exponential decay.

These results indicate that the time evolution of the maximum cluster size does not follow a simple power-law relaxation but is more appropriately characterized as an exponential decay process. 
A single exponential function cannot describe the relaxation. Here, we propose that the superposition of two exponential functions provides a better description. Specifically, as $\phi_{\mathrm{max}}(t)$ approaches the saturation value $\phi_{\mathrm{max}}(\infty)$, two distinct stages, which are one relatively rapid initial relaxation and a slower subsequent relaxation, are observed and can be approximated by the following expression:
\begin{equation}
  \phi_{\mathrm{max}}(t)
  = \phi_{\mathrm{max}}(\infty) - A_1 \exp\Bigl(-\frac{t}{\tau_1}\Bigr) - A_2 \exp\Bigl(-\frac{t}{\tau_2}\Bigr),
  \label{eq:double_exponential_fit}
\end{equation}
where $\tau_1$ and $\tau_2$ are the relaxation time scales corresponding to the initial and later relaxation stages, respectively, and $A_1$ and $A_2$ are amplitude constants. 
Equation~\eqref{eq:double_exponential_fit} encapsulates the two-stage relaxation behavior that cannot be captured by a single exponential function and is consistent with the observation in Fig.~\ref{fig:maxcluster_timeevo_L3}(b), where two distinct linear regions are apparent. 
Although the relaxation time scales $\tau_1$ and $\tau_2$ are found to depend on the system size, the precise nature of this dependence remains a subject for future investigation.

\begin{figure}
\centering
\includegraphics[width=\linewidth]{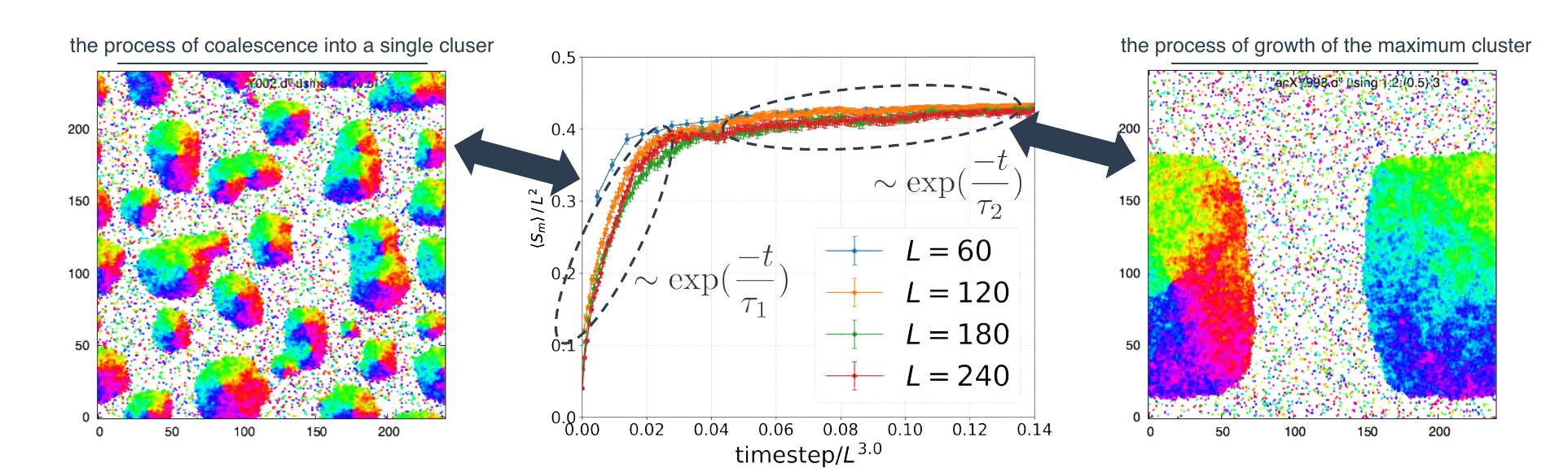}
\caption{(Color online) Two exponential relaxation processes and snapshots of corresponding times. In rapid relaxation, the dynamics of small clusters are dominated. In contrast, the latter relaxation is caused by the growth of a single large cluster.}	
\label{fig:exp_explain}
\end{figure}
The following physical mechanisms can explain the two-stage relaxation observed; 
The initial rapid relaxation with the $\tau_1$ relaxation time scale is thought to correspond to the stage in which small clusters and isolated particles scattered throughout the system collide and merge rapidly over short distances to form moderately large clusters. 
In contrast, the latter relaxation with the $\tau_2$ relaxation time scale corresponds to a much slower process in which a single large cluster, formed by repeated coalescence and aggregation, gradually grows to its maximum size by slowly absorbing diffusing particles. Snapshots of the system corresponding to two relaxation regimes are shown in Fig.~\ref{fig:exp_explain}.
This behavior is also analogous to the dynamics of the order parameter in first-order phase transitions, characterized by the nucleation of multiple clusters, coarsening, and subsequent growth of these clusters. 

\begin{figure}
\centering
\subfigure[$\rho=0.3$]{\includegraphics[width=0.48\linewidth]{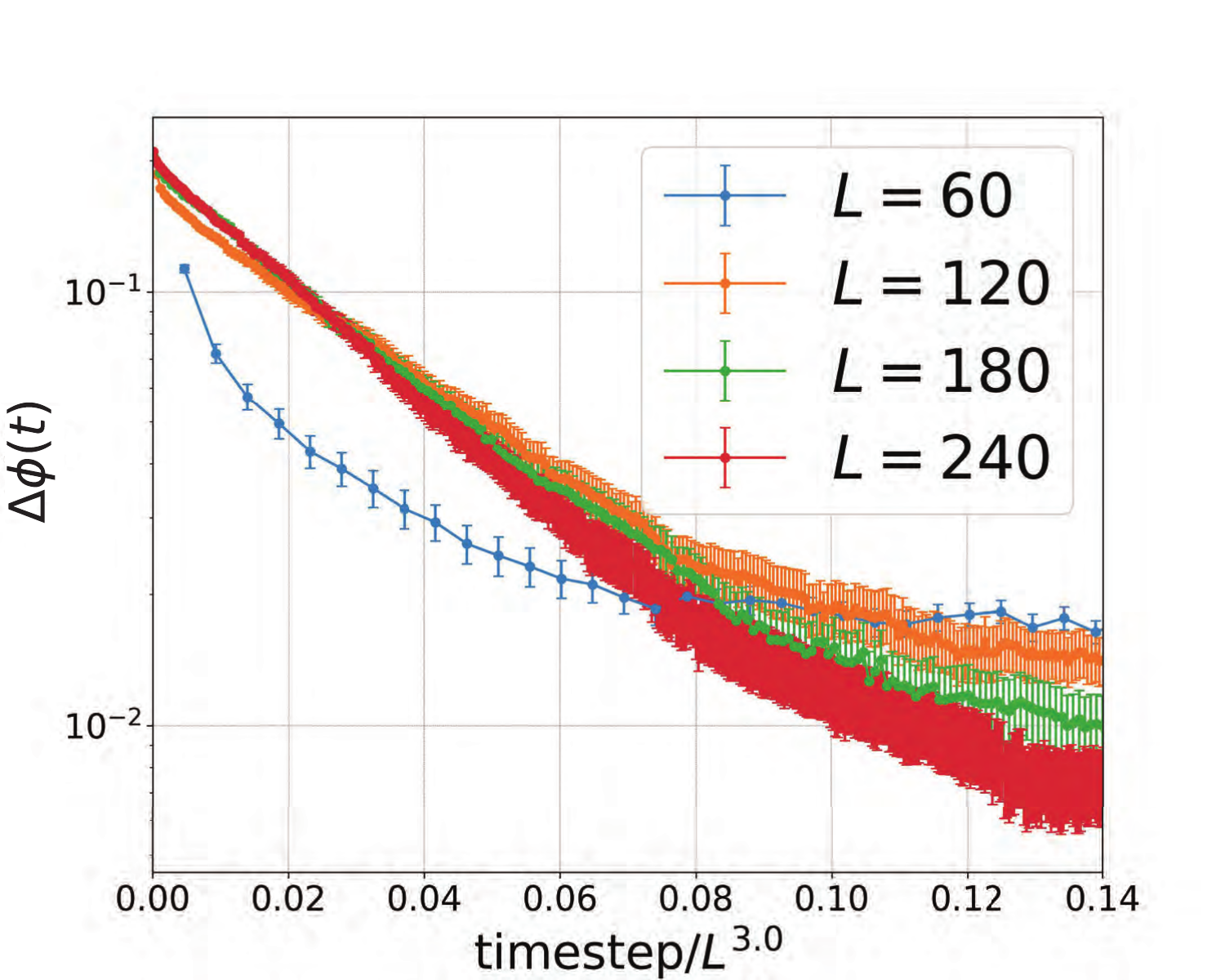}}
\subfigure[$\rho=0.4$]{\includegraphics[width=0.48\linewidth]{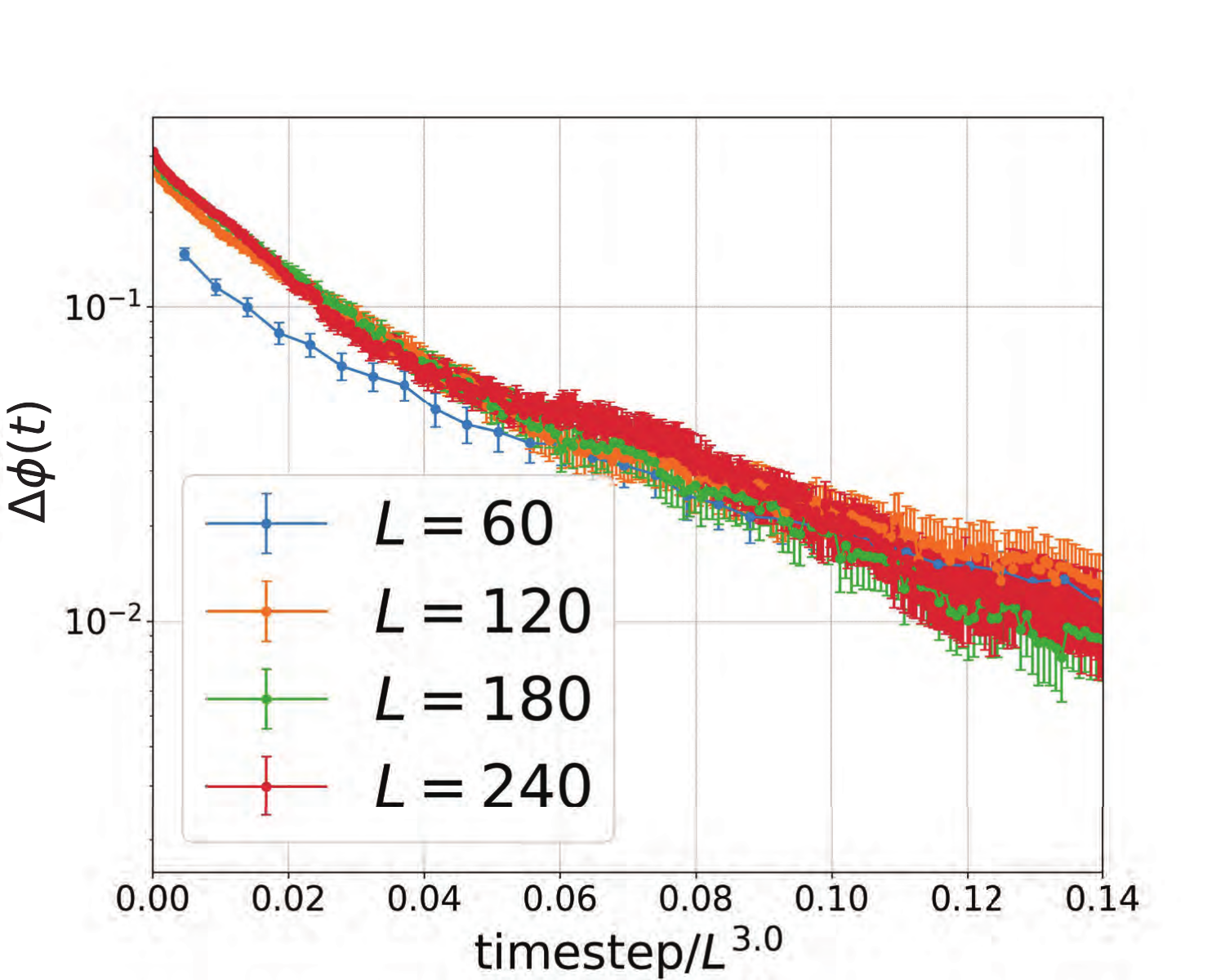}}  
\subfigure[$\rho=0.5$]{\includegraphics[width=0.48\linewidth]{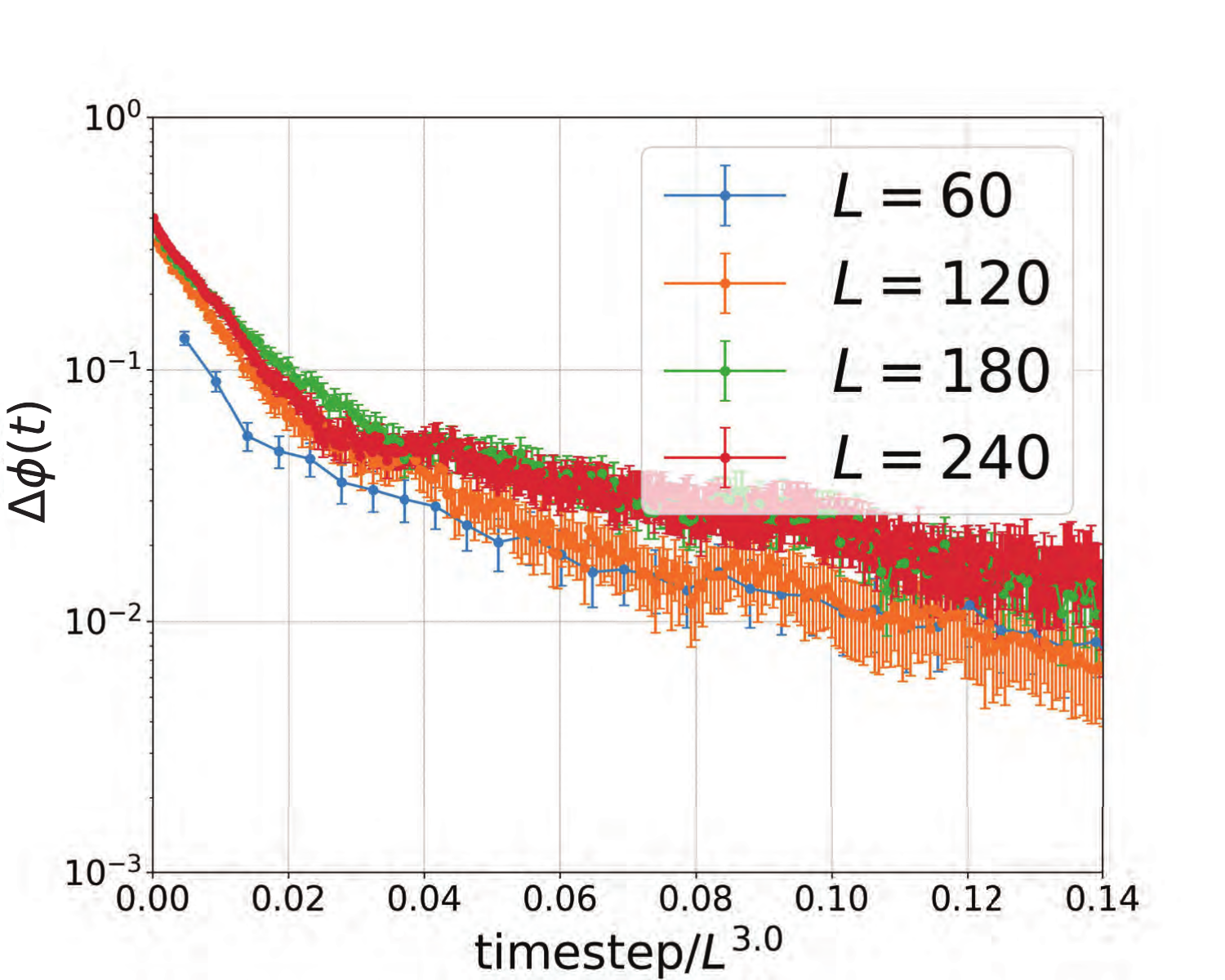}}
\subfigure[$\rho=0.6$]{\includegraphics[width=0.48\linewidth]{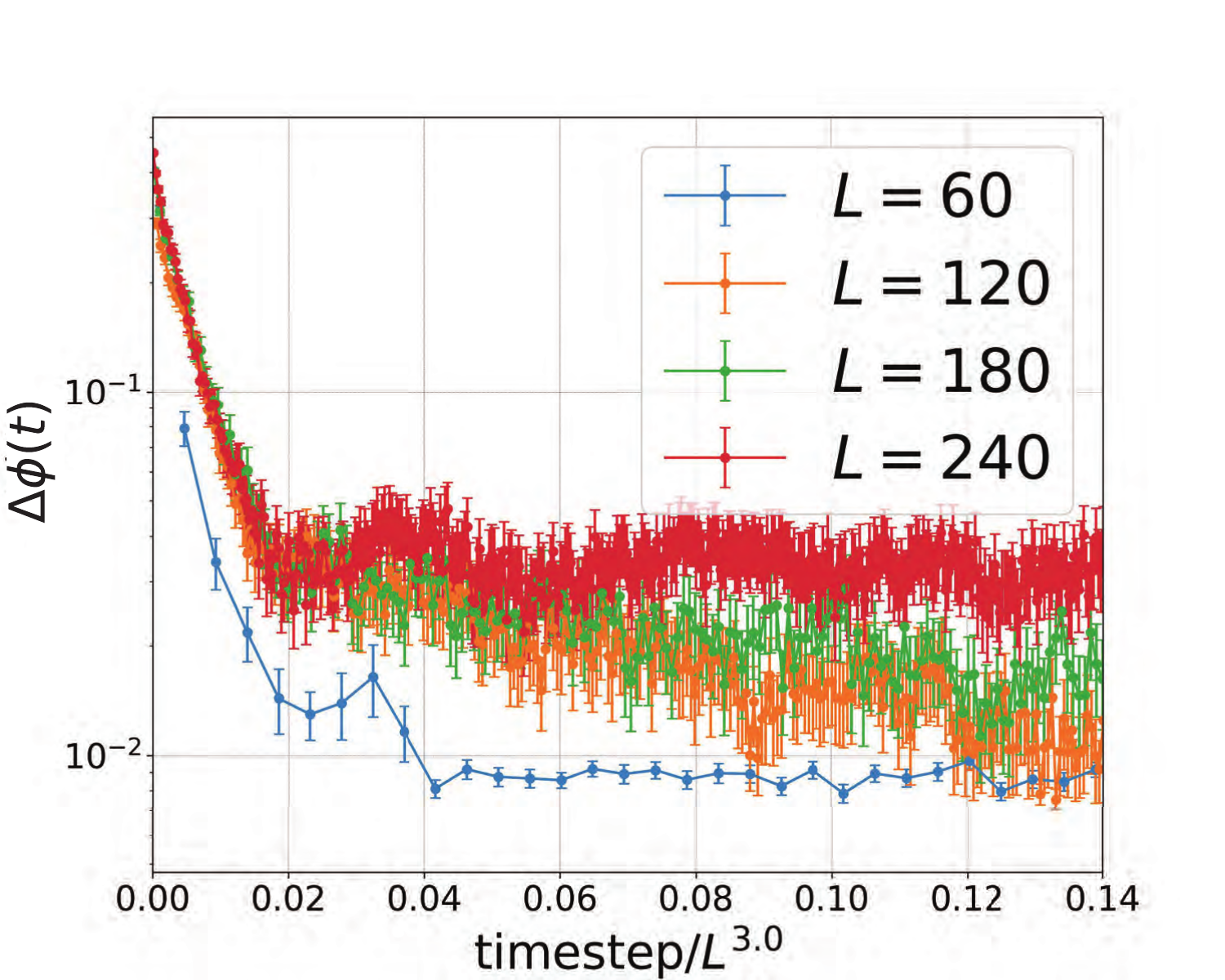}}
\caption{(Color online) Time evolution of the area fraction of the maximum cluster size $\phi_{\mathrm{max}}(t)$, shown on a semi-logarithmic plot, for system sizes $L=60$, $120$, $180$, and $240$ under different density conditions: (a) $\rho=0.3$, (b) $\rho=0.4$, (c) $\rho=0.5$, and (d) $\rho=0.6$. In all cases, the data indicate that the growth is described as the sum of two exponential functions. Interestingly, when the time axis is normalized by $L^3$, the transition between the two exponential regimes occurs over a similar time interval, independent of the system size.}
\label{fig:cluster_timestep_scale_exp}
\end{figure}
Finally, we compare the two-stage relaxation for the case of other densities $\rho=0.3$, $\rho=0.4$, and $\rho=0.6$ in Fig.~\ref{fig:cluster_timestep_scale_exp}. 
In this figure, the time evolution of the area fraction of the maximum cluster size $\phi_{\mathrm{max}}(t)$ is plotted on a semi-logarithmic scale for different system sizes, $L=60$, $120$, $180$, and $240$, under each density condition. 
The figure shows that the data can be described as the sum of two exponential functions regardless of density. 
Interestingly, when the time axis is scaled by $L^3$, the crossover between the two exponential regimes occurs over a similar scaled time interval for all system sizes. This observation further supports the notion that the characteristic time scale for cluster growth is $\tau\sim L^3$.

\section{Summary and Discussion}

In this study, we constructed a self-propelled lattice-gas active XY (SPLG-AXY) model that integrates the features of the classical XY and Vicsek models. 
Its computational performance is more efficient than that of the Vicsek and AMB+ models.
This, therefore, allows numerical studies of a larger parameter space. Furthermore, because it is a lattice model, parallel computation is also possible by modifying its update rules.
In this article, we conducted numerical simulations varying particle density and self-propulsion strength. The main conclusions obtained in this work are summarized below.

First, we found a correlation between topological defects and particle aggregation.
In the high self-propulsion regime, particles aggregate around the topological defects with $m=+1$ vortex charge, facilitating cluster formation.
In contrast, all topological defects with $m=-1$ charge vanish from the system. 
It was confirmed that as the self-propulsion strength $\epsilon$ increases, the average total vorticity, defined as the average of the difference of the numbers of the topological defects with $m=+1$ and $m=-1$ charges, becomes positive except for the high density. 
Furthermore, comparing the total vorticity heatmap with the cluster distribution snapshots reveals a strong correlation between topological defects and motility-induced phase separation. 
Even a small self-propulsion is sufficient to induce phase separation, particularly at high particle densities.

Next, we analyzed the time evolution of the cluster size distribution and its dependence on system size. 
Analysis of the cumulative distribution of cluster sizes for varying system sizes revealed that a single giant cluster eventually forms in a steady state. Larger systems, however, require a longer relaxation time to reach a steady state. 
Moreover, the area fraction of the maximum cluster size is found to be independent of the system size and is determined solely by the particle density $\rho$.

In addition, we analyzed the time evolution of the area fraction of the largest cluster. Then, we found that there is a two-stage exponential relaxation and scaling of its characteristic time; 
By analyzing the formation process of the giant cluster, we observed that, when time is scaled by $L^3$, data from different system sizes collapse onto a single curve. 
This collapse suggests that the characteristic time scale for cluster growth follows a scaling law of the form $\tau\sim L^3$.
Furthermore, semi-logarithmic plotting of the time evolution of the maximum cluster size reveals that relaxation can be described as the superposition of two exponential functions: A rapid initial relaxation corresponding to the merging of small clusters and a slower subsequent relaxation corresponding to the coalescence of an enlarged cluster.

During the study, we found that the phase separation behavior is similar to that observed in the coexistence phase of a first-order phase transition. In the coexistence phase of the first-order phase transition, a spatial characteristic scale $\xi$ of the conserved order parameter cluster shows a power-law growth $\xi \sim t^{1/3}$. 
Despite being of a nonequilibrium nature, several active-matter models also exhibit this behavior.\cite{Cates2015,thompson2011,adachi2020universality,stenhammar2013continuum}
There is, however, a slight difference between the present model and other phase-separation models, whether in equilibrium or non-equilibrium.
In the SPLG-AXY model, this exponent is observed in the relaxation-time scaling with the system size $L$ as $\tau\sim L^3$. 
At least with respect to the final stage of the relaxation, this is a finite-size effect.
If we regard $\xi$ as $L$ because only a single cluster extends to the system size in the final stage of relaxation, we can expect the scaling $\tau \sim L^{3}$ from the power-law cluster growth $\xi \sim t^{1/3}$. 
However, the physical reason why the relaxation time scales as $L^{3}$, and why the early-stage relaxation is scaled by the same scale, remains unclear.

The interplay between the topological defects inherited from the classical XY model and the exclusion effects of self-propelled particles leads to diverse aggregation patterns around these defects. 
These findings imply that topological defects play a significant role in active matter systems and that the analogy with equilibrium phase separation can enhance our understanding of these phenomena. 
An important aspect not addressed in this study is the behavior of the spin-particle correlation function in the SPLG-AXY model. 
The contrasting ordering behaviors, that is, quasi-long-range order in the classical XY model and long-range order in the Vicsek model, make the correlation function dynamics of this intermediate model particularly worthy of investigation.

In summary, our study using the SPLG-AXY model has advanced the understanding of the impact of topological defects and the dynamics of cluster formation in active matter systems.
The strong coupling between topological defects and particle aggregation, and the observation that cluster growth can be characterized by exponential relaxation and scaling laws, are important findings for drawing analogies between equilibrium and active systems. 
The present model, inspired by the Vicsek model, was constructed on the lattice. 
This model resembles the Toner-Tu model, a continuous version of the Vicsek model. 
But the analysis shows that the present model is similar to the AMB+ model, because it lacks a traveling-wave phase. This is due to the diffusive self-propulsion rule. 
Recent studies on the AMB and AMB+ models have confirmed that the domain growth law exhibits a crossover from $t^{1/3}$ to $t^{1/4}$. \cite{PMP21,YMP25}
Whether analogous behavior exists in the SPLG-AXY model remains an open question and constitutes a promising direction for future research.

\begin{acknowledgment}
The authors are grateful to members of the Hatano lab for valuable discussions.
This work was partly supported by JSPS KAKENHI Grant Numbers 19K03652 and 24K06899.
\end{acknowledgment}

\appendix

\section{Computational Performance}
\label{sec:compperf}

This section summarizes the computational performance of the SPLG-AXY model. 
First, we summarize benchmarks compared with other models. 
Benchmarks were performed on a single thread on the Apple M4 Pro processor of the Apple Mac Mini using the Apple Clang compiler version 17.0.0 with the compile option ``-O3 -ffast-math''.
The result is compared with the Vicsek model implemented with the cell method \cite{allen2017} and the active model B plus (AMB+) \cite{NFTWTC17,TNC2018}, which is a continuum field-theoretical model, spatially discretized for simulation and using the Euler scheme for time integration.

The computational conditions are as follows: the system is two-dimensional, and the linear system sizes are $L=100, 200$, and $400$, corresponding to the lattice sizes for the SPLG-AXY and AMB+ models, and to the system length for the Vicsek model. The periodic boundary condition is imposed. 
Number densities are examined with $\rho=0.25, 0.5,$ and $0.75$ for the SPLG-AXY model and the Vicsek model. In the AMB+ model, the number density has no mean due to the continuum field-theoretical model. For the above models, we calculate for $t = 10 \, 000$ steps from the initial conditions. 
It should be mentioned that the SPLG-AXY model and the Vicsek model are discrete-time map models. Thus, the above time represents the actual time. But the AMB+ model is continuous-time. Therefore, the actual time of the AMB+ model, compared with the SPLG-AXY model and the Vicsek model, should be multiplied by a small time step $\Delta t$.

The model parameters are $T=1/4$ and $\epsilon = 1$ for the SPLG-AXY model, $v=0.5, \eta=0.3, $and $r=1$ for the Vicsek model, whose symbols are defined in the original paper \cite{vicsek1995novel}, and $A=0.25, K=1, K_{1}=0, D=0.2, \zeta =4, \lambda=1, \Delta x=1, \Delta t=0.01$ for the AMB+ model, whose symbols are defined in the paper \cite{TNC2018}.

\begin{figure}
\centering
\includegraphics[width=0.48\linewidth]{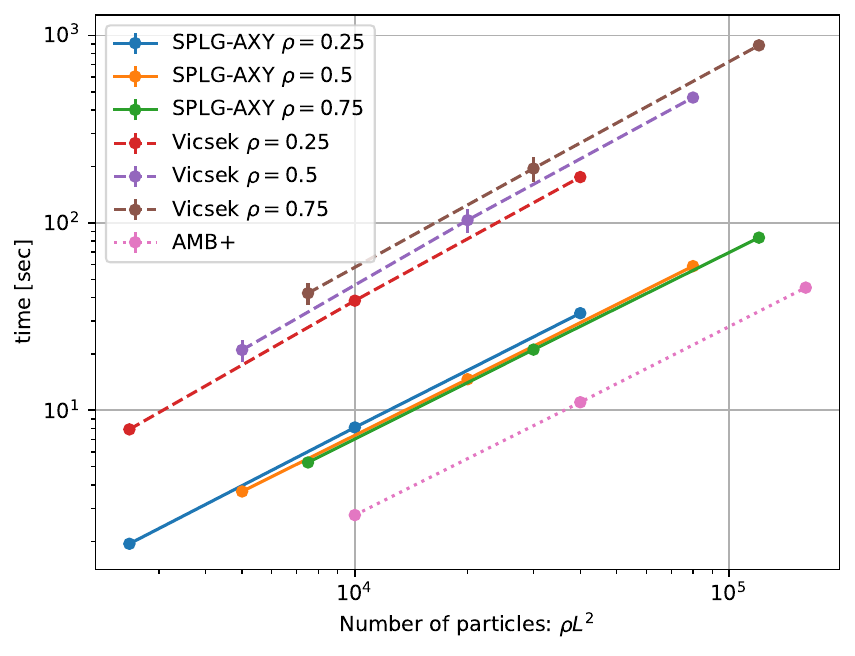}
\caption{(Color online) Benchmark results of the SPLG-AXY model compared with the Vicsek model and the active model B plus. The elapsed time is averaged over $10$ samples, and error bars show the 
square root of the variance.}
\label{fig:benchmark}
\end{figure}
The results are shown in Fig.~\ref{fig:benchmark}. The horizontal axis represents the number of particles in the SPLG-AXY and Vicsek models, and the number of lattice sites in the AMB+ model. 
The vertical axis shows the elapsed time for the above steps. 
The elapsed time is averaged over $10$ samples, and error bars show the square root of the variance.
This result indicates that all models exhibit linear scaling with respect to the number of degrees of freedom. Additionally, the Vicsek model shows a density-dependent increasing prefactor in the scaling. This is because the number of interacting particles increases with density. The SLPG-AXY model shows almost no such dependency. The model with the shortest computation time is the AMB+ model; however, since it is a continuous-time model, to align the characteristic time scales across all models, it is necessary to multiply the horizontal axis of the AMB+ model by $\Delta t$. Thus, the SPLG-AXY model has the highest computational efficiency.

\begin{figure}
\centering
\includegraphics[width=0.48\linewidth]{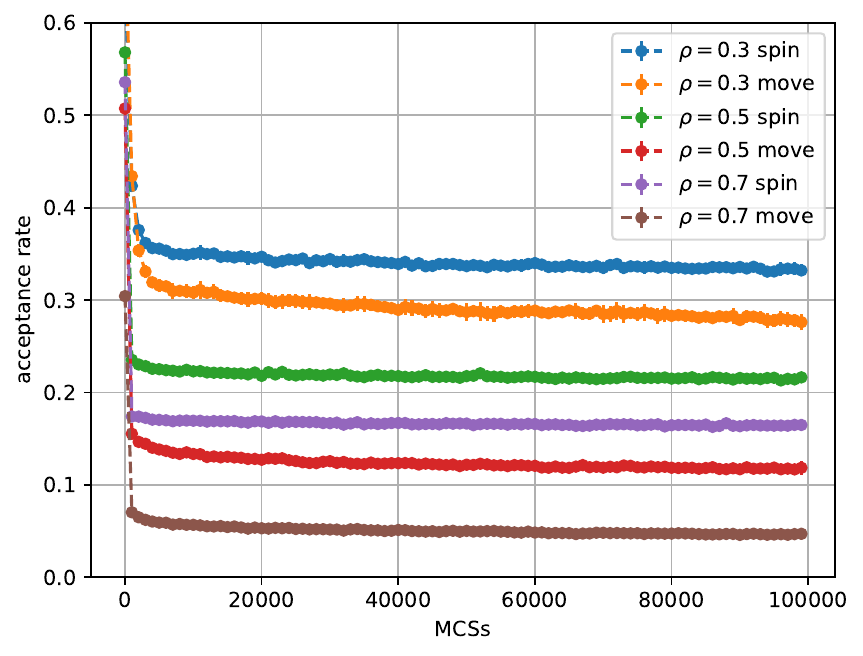}
\caption{(Color online) Acceptance rates of the SPLG-AXY model. There are two types of acceptance rates, spin and move updates. The acceptance rates are averaged over $10$ samples, and error bars show the square root of the variance. The system size is $L=200$, and the density is $\rho=0.3, 0.5$ and $0.7$.}
\label{fig:acceptance}
\end{figure}
Next, we calculate the acceptance rate per Monte Carlo updates in the SPLG-AXY model.
There are two types of acceptance rates, spin and move updates. The rates are dependent on the density $\rho$ and time. Thus, we calculate two rates as a time series from the random initial state and show them in Fig.~\ref{fig:acceptance}. 
For examined densities, each spin acceptance rate is higher than the move acceptance rate, and as the density increases, the acceptance rates decrease. 
As time evolves, the MIPS structure develops. Thus, the acceptance rates decrease.


\begin{thebibliography}{10}

\bibitem{Marchetti2013}
M.~C. Marchetti, J.~F. Joanny, S.~Ramaswamy, T.~B. Liverpool, J.~Prost, M.~Rao,
  and R.~A. Simha: Rev. Mod. Phys. \textbf{85}, 1143 (2013).

\bibitem{Bechinger2016}
C.~Bechinger, R.~Di~Leonardo, H.~L{\"o}wen, C.~Reichhardt, G.~Volpe, and
  G.~Volpe: Rev. Mod. Phys. \textbf{88}, 045006 (2016).

\bibitem{Toner2024}
J.~L. Toner: {\em The Physics of Flocking: Birth, Death, and Flight in Active Matter} (Cambridge University Press, Cambridge, 2024).

\bibitem{Reynolds1987}
C.~W. Reynolds: ACM SIGGRAPH Comput. Graph. \textbf{21}, 25 (1987).

\bibitem{Shimoyama1996}
N.~Shimoyama, K.~Sugawara, T.~Mizuguchi, Y.~Hayakawa, and M.~Sano: Phys.
  Rev. Lett. \textbf{76}, 3870 (1996).

\bibitem{niwaNewtonianDynamicalApproach1996}
H.-S. Niwa: J. Theor. Biol. \textbf{181}, 47 (1996).

\bibitem{ItoUchida2022}
S.~Ito and N.~Uchida: Europhys. Lett. \textbf{138}, 17001 (2022).

\bibitem{itoEmergenceGiantRotating2022}
S.~Ito and N.~Uchida: J. Phys. Soc. Jpn. \textbf{91}, 
  064806 (2022).

\bibitem{kumar2010}
M.~S. Kumar and P.~Philominathan: Biophys. Rev. \textbf{2}, 13 (2010).

\bibitem{nishiguchi2025}
D.~Nishiguchi, S.~Shiratani, K.~A. Takeuchi, and I.~S. Aranson: Proc.
  Natl. Acad. Sci. USA. \textbf{122}, e2414446122 (2025).

\bibitem{kawaguchi2017topological}
K.~Kawaguchi, R.~Kageyama, and M.~Sano: Nature \textbf{545}, 327 (2017).

\bibitem{sumino2012large}
Y.~Sumino, K.~H. Nagai, Y.~Shitaka, D.~Tanaka, K.~Yoshikawa, H.~Chat{\'e}, and
  K.~Oiwa: Nature \textbf{483}, 448 (2012).

\bibitem{vicsek1995novel}
T.~Vicsek, A.~Czir{\'o}k, E.~Ben-Jacob, I.~Cohen, and O.~Shochet: Phys.
  Rev. Lett. \textbf{75}, 1226 (1995).

\bibitem{stanley1968dependence}
H.~E. Stanley: Phys. Rev. Lett. \textbf{20}, 589 (1968).

\bibitem{tobochnik1979monte}
J.~Tobochnik and G.~Chester: Phys. Rev. B \textbf{20}, 3761 (1979).

\bibitem{kosterlitz1973}
J.~M. Kosterlitz and D.~J. Thouless: J. Phys. C: Solid State Phys.
  \textbf{6}, 1181 (1973).

\bibitem{kosterlitz1974critical}
J.~M. Kosterlitz: J. Phys. C: Solid State Phys. \textbf{7}, 1046 (1974).

\bibitem{berezinskii1971destruction}
V.~L. Berezinskii: Sov. Phys. JETP \textbf{32}, 493 (1971).

\bibitem{berezinskii1972destruction}
V.~L. Berezinskii: Sov. Phys. JETP \textbf{34}, 610 (1972).

\bibitem{Nelson2002}
D.~R. Nelson: {\em Defects and Geometry in Condensed Matter Physics} (Cambridge
  University Press, Cambridge, 2002).

\bibitem{shankar2022}
S.~Shankar, A.~Souslov, M.~J. Bowick, M.~C. Marchetti, and V.~Vitelli: Nat.
  Rev. Phys. \textbf{4}, 380 (2022).

\bibitem{Cates2015}
M.~E.~Cates and J.~Tailleur: Annu. Rev. Condens. Matter Phys.
  \textbf{6}, 219 (2015).

\bibitem{ST13} A.~P.~Solon and J.~Tailleur: Phys. Rev. Lett. \textbf{111},  078101 (2013).

\bibitem{kourbane-houssene2018}
M.~{Kourbane-Houssene}, C.~Erignoux, T.~Bodineau, and J.~Tailleur: Phys.
  Rev. Lett. \textbf{120}, 268003 (2018).

\bibitem{AadachiNakano2024}
K.~Adachi and H.~Nakano: Phys. Rev. Res. \textbf{6}, 033234 (2024).

\bibitem{NakanoAdachi2024}
H.~Nakano and K.~Adachi: Phys. Rev. Res. \textbf{6}, 013074 (2024).

\bibitem{toner1995long}
J.~Toner and Y.~Tu: Phys. Rev. Lett. \textbf{75}, 4326 (1995).

\bibitem{tonerFlocksHerdsSchools1998}
J.~Toner and Y.~Tu: Phys. Rev. E \textbf{58}, 4828 (1998).

\bibitem{Toner2012}
J.~Toner: Phys. Rev. E \textbf{86}, 031918 (2012).

\bibitem{NFTWTC17}
C. Nardini, É. Fodor, E. Tjhung, F. Van Wijland, J. Tailleur, and M.~E. Cates, Phys. Rev. X \textbf{7}, 021007 (2017).


\bibitem{TNC2018}
E. Tjhung, C. Nardini, and M.~E. Cates, Phys. Rev. X \textbf{8}, 031080 (2018).

\bibitem{haldarMobilityinducedOrderActive2023}
A.~Haldar, A.~Sarkar, S.~Chatterjee, and A.~Basu: Phys. Rev. E \textbf{108}, L032101 (2023).

\bibitem{haldarActiveXYModel2023}
A.~Haldar, A.~Sarkar, S.~Chatterjee, and A.~Basu: Phys. Rev. E \textbf{108}, 034114 (2023).

\bibitem{fisher1963statistical}
R.~A. Fisher and F.~Yates: {\em Statistical tables for biological, agricultural and medical research} (Edinburgh: Oliver and Boyd, 1963).

\bibitem{knuth1997}
D.~E. Knuth: {\em The Art of Computer Programming, volume 2, Seminumerical Algorithms} (Addison-Wesley, Reading, Massachusetts, 1998) 3rd ed.

\bibitem{hoshen1976percolation}
J.~Hoshen and R.~Kopelman: Phys. Rev. B \textbf{14}, 3438 (1976).

\bibitem{LS1961a}
I.~Lifshitz and V.~Slyozov: J. Phys. Chem.  Solids  \textbf{19}, 35 (1961).

\bibitem{CL1995}
P.~M. Chaikin and T.~C. Lubensky: {\em Principles of Condensed Matter Physics}
  (Cambridge University Press, 1995).

\bibitem{thompson2011}
A.~G. Thompson, J.~Tailleur, M.~E. Cates, and R.~A. Blythe: J.
  Stat. Mech.: Theory Exp. \textbf{2011}, P02029 (2011).

\bibitem{adachi2020universality}
K.~Adachi and K.~Kawaguchi: arXiv:2012.02517  (2020).

\bibitem{stenhammar2013continuum}
J.~Stenhammar, A.~Tiribocchi, R.~J. Allen, D.~Marenduzzo, and M.~E. Cates:
  Phys. Rev. Lett. \textbf{111}, 145702 (2013).

\bibitem{PMP21}
S.~Pattanayak, S.~Mishra, and S.~Puri: Phys. Rev. E \textbf{104}, 014606 (2021).

\bibitem{YMP25}
P.~K.~Yadav, S.~Mishra, and S.~Puri: Phys. Rev. E \textbf{112}, 035412 (2025).

\bibitem{allen2017}
M.~P. Allen and D.~J. Tildesley: {\em Computer Simulation of Liquids}
(Oxford University Press, Oxford, United Kingdom, 2017) 2nd ed.
\end{thebibliography}
\end{document}